\documentclass[11pt,a4paper]{article}
\pdfoutput=1
\usepackage{jcapmod}
\usepackage{shorthand}

\usepackage{mathtools}
\usepackage{booktabs}
\usepackage[english]{babel}
\usepackage{amsmath,amssymb,amsbsy,amstext, amsthm, simplewick, amsfonts}
\usepackage{hyperref}
\usepackage{comment}
\usepackage{graphicx}
\usepackage[small]{caption}
\usepackage{siunitx}
\usepackage{upgreek}
\usepackage{framed}
\usepackage{wrapfig}
\usepackage{multirow}
\usepackage{bbm}
\usepackage[svgnames,dvipsnames,x11names]{xcolor}
\usepackage[utf8x]{inputenc}
\usepackage{selinput}
\usepackage{bm}
\usepackage{float}
\usepackage{geometry}
\usepackage{yfonts}
\usepackage{caption}
\usepackage{subcaption}
\usepackage{sidecap}
\usepackage{longtable}
\usepackage{anyfontsize}
\usepackage{dsfont}
\usepackage[normalem]{ulem}

\usepackage[framemethod=default]{mdframed}
\newmdenv[skipabove=7pt,
skipbelow=7pt,
rightline=false,
leftline=false,
topline=false,
bottomline=false,
backgroundcolor=gray!10,
linecolor=gray,
innerleftmargin=5pt,
innerrightmargin=5pt,
innertopmargin=5pt,
innerbottommargin=5pt,
leftmargin=0cm,
rightmargin=0cm,
linewidth=4pt]{eBox}
\newmdenv[skipabove=7pt,
skipbelow=7pt,
rightline=true,
leftline=true,
topline=true,
bottomline=true,
backgroundcolor=white,
linecolor=gray,
innerleftmargin=5pt,
innerrightmargin=5pt,
innertopmargin=5pt,
innerbottommargin=5pt,
leftmargin=0cm,
rightmargin=0cm,
linewidth=1pt]{eBox2}

\definecolor{blue3}{RGB}{31, 119, 180}
\definecolor{red3}{RGB}{	214, 39, 40}
\definecolor{orange3}{RGB}{255, 127, 14}
\definecolor{green3}{RGB}{44, 160, 44}
\definecolor{repBlue}{RGB}{31, 119, 180}
\definecolor{repRed}{RGB}{	214, 39, 40}
\definecolor{repGreen}{RGB}{44, 160, 44}

\renewcommand{\(}{\left(}
\renewcommand{\)}{\right)}
\renewcommand{\[}{\left[}
\renewcommand{\]}{\right]}

\def\be{\begin{equation}}
\def\ee{\end{equation}}
\newcommand{\bea}{\begin{eqnarray}}
\newcommand{\eea}{\end{eqnarray}}
\def\fnl{f_{\rm NL}}

\usepackage{colortbl}
\definecolor{lightgreen}{cmyk}{0.2, 0, 0.2, 0.2}
\definecolor{lightgray}{cmyk}{0.1,0.2,0,0.1}
\definecolor{lightgray2}{cmyk}{0.1,0.1,0,0.1}

\makeatletter
\newlength{\apb@width}
\newcommand{\autoparbox}[2][c]{\settowidth{\apb@width}{#2}\parbox[#1]{\apb@width}{#2}}

\makeatother


\def\beq{\begin{equation}}
\def\eeq{\end{equation}}

\newcommand{\fNL}{{f_\mathrm{NL}}}
\newcommand{\ells}{{\ell_1 \ell_2 \ell_3}}

\allowdisplaybreaks[1]
\begin{document}

\newgeometry{top=2cm, bottom=2cm, left=2.9cm, right=2.9cm}

\begin{titlepage}
\setcounter{page}{1} \baselineskip=15.5pt 
\thispagestyle{empty}

\begin{center}
{\fontsize{14.4}{18} \bf Searching for  Cosmological Collider in the Planck CMB Data \\ [16pt]
\fontsize{16}{18} \it II: collider templates and Modal analysis}\\
\end{center}

\vskip 16pt

\begin{center}
\noindent
{\fontsize{12}{18}\selectfont Petar Suman$^1$, Dong-Gang Wang$^{2,1}$, Wuhyun Sohn$^3$, \\ [8pt] James R. Fergusson$^1$ and  E. P. S. Shellard$^1$}
\end{center}

\vskip 16pt

\begin{center}

  \vskip8pt
{$^1$\fontsize{10.2}{18}\it Centre for Theoretical Cosmology, Department of Applied Mathematics and Theoretical Physics\\ University of Cambridge,
Wilberforce Road, Cambridge, CB3 0WA, UK}

  \vskip8pt
{$^2$ \fontsize{10.2}{18}\it Institute for Advanced Study and Department of Physics,\\ Hong Kong University of Science and Technology, Clear Water Bay, HK, China
}

  \vskip8pt
{$^3$ \fontsize{10.2}{18}\it AstroParticule et Cosmologie 
10, rue Alice Domon et Léonie Duquet, 75013 Paris, France
}

\end{center}

\vskip 20pt

\vspace{0.4cm}
 \begin{center}{\bf Abstract} 
 \end{center}
 \noindent
Signatures of massive particles during inflation are highly informative targets for cosmological experiments. 
With recent progress on both theoretical and observational frontiers, we have reached the point where these novel signals of primordial non-Gaussianities (PNG) can be systematically tested with increasingly precise data.
In this paper, we present the results of improved CMB data analysis for cosmological collider signals using Planck CMB data.
To set the stage, we first construct a set of simplified but characteristic collider templates which are accurate over a broad range of particle masses, spins and sound speeds. In order to break degeneracies with single-field PNG, we propose an orthogonalization scheme such that the collider templates are uncorrelated with the highly constrained equilateral and orthogonal shapes. 
On this basis, we deploy the Modal bispectrum estimator for the Planck analysis and perform a systematic scan of parameters to search for the most significant collider signal.
The maximum signal-to-noise ratio is found to be $2.35\sigma$ for an orthogonalized massive spin-0 exchange after taking into account the look-elsewhere effect.
In addition, we cross-validate the Modal analysis with the CMB-BEST pipeline, which demonstrates the consistency of results across the benchmark examples of collider templates.
Given the low signal-to-noise ratio regime we find at the current stage of PNG observations, we believe 
the orthogonalization procedure provides an optimized strategy for future tests of the cosmological collider signals.

\noindent

\end{titlepage}

\newpage

\restoregeometry
\setcounter{tocdepth}{3}
\setcounter{page}{1}
\tableofcontents

\vskip 16pt

 \hrule

\vskip 25pt

\section{Introduction}
\label{sec:intro}

Modern advances in cosmology provide an unsurpassed opportunity for probing fundamental physics. Cosmic inflation, which may be the highest-energy event accessible in nature, serves as a natural laboratory at extremely high energies. 
Physical processes during inflation can leave observable imprints in today's cosmological data, including measurements of the Cosmic Microwave Background (CMB) and surveys of Large Scale Structure (LSS).
As the characteristic energy scale of inflation can be as high as $10^{13}$GeV, far beyond the reach of the terrestrial collider experiment, it is therefore interesting to study precise predictions of inflation as a test of unknown particle physics in the microscopic regime and even aspects of quantum gravity \cite{Meerburg:2019qqi,Achucarro:2022qrl}.

Among many theoretical proposals, the cosmological collider scenario has emerged as a particularly compelling and model-independent approach to probing new physics during inflation.
Drawing an analogy with conventional particle colliders, this program exploits the fact that massive particles present during inflation can leave characteristic imprints in the correlation functions of primordial fluctuations \cite{Chen:2009zp,Baumann:2011nk,Noumi:2012vr,Arkani-Hamed:2015bza}. In principle, precise measurements of these features would allow one to infer both the mass spectrum and spin content of particles present during inflation, providing direct insight into physics at ultra-high energies.
This prospect has motivated extensive theoretical efforts over the past decade \cite{Chen:2009we, Assassi:2012zq, Chen:2012ge, Pi:2012gf,  Chen:2015lza, Lee:2016vti, Chen:2016uwp, Chen:2016hrz, Chen:2017ryl, Kehagias:2017cym, Kumar:2017ecc, An:2017hlx, An:2017rwo, Baumann:2017jvh, Chen:2018xck, Kumar:2018jxz, Bordin:2018pca,  Kim:2019wjo, Alexander:2019vtb,  Wang:2019gbi, Wang:2019gok, Aoki:2020zbj, Lu:2021wxu, Wang:2021qez, Tong:2021wai, Cui:2021iie, Tong:2022cdz, Chen:2022vzh}.
In particular, significant progress has been made in computing cosmological collider signals using a variety of analytic techniques.
Notably, modern approaches, such as the cosmological bootstrap, have enabled the systematic construction of full bispectrum shapes consistent with fundamental principles such as unitarity, locality, and spacetime symmetries.
These developments have clarified the generic features of collider signals and expanded the range of theoretically controlled templates available for data analysis \cite{Arkani-Hamed:2018kmz,Baumann:2019oyu,Baumann:2020dch, Arkani-Hamed:2017fdk, Benincasa:2018ssx, Sleight:2019mgd, Sleight:2019hfp, Goodhew:2020hob, Cespedes:2020xqq, Pajer:2020wxk, Pimentel:2022fsc, Jazayeri:2022kjy, Baumann:2022jpr, Qin:2022fbv, Xianyu:2022jwk, Wang:2022eop,Qin:2023ejc,Stefanyszyn:2023qov, Jazayeri:2023xcj, Aoki:2024uyi, McCulloch:2024hiz,Stefanyszyn:2024msm,Liu:2024xyi, Pajer:2024ckd, Cespedes:2025dnq, Pimentel:2025rds, Wang:2025qww, Wang:2025qfh, Qin:2025xct, deRham:2025mjh, Stefanyszyn:2025yhq, Kumar:2025anx,Cespedes:2025ple,Colas:2025ind,Jazayeri:2025vlv,Cheung:2025dmc,Xianyu:2025lbk}.

On the observational side, however, the search for primordial non-Gaussianities (PNG) remains a challenge.
Most efforts to date have focused on the three simplest and most commonly studied templates: local, equilateral, and orthogonal shapes of the primordial bispectrum. 
The \textit{Planck} satellite has placed stringent constraints on the amplitudes of these shapes \cite{Akrami:2019izv}, finding them consistent with zero.
This indicates that we are in the low signal-to-noise regime for the search of PNG, which may lead to the pessimistic view that the detection of PNG is beyond the reach of our current and forthcoming cosmological experiments.
Nevertheless, tests of PNG are {\it highly template-dependent} -- feeding different bispectrum shapes into the data analysis can yield totally distinct results of observational constraints.
Therefore, identifying theoretically well-motivated shapes and constructing accurate templates is essential for any systematic search of PNG.

Most notably, the recent progress in the cosmological collider program provides a full set of bispectrum predictions that significantly enlarge our library of PNG templates.
This has motivated recent dedicated searches for cosmological colliders in both CMB \cite{Sohn:2024se} and LSS \cite{Cabass:2024wob} datasets.
For the CMB analysis, using \textit{Planck} data, the maximum signal-to-noise ratio identified so far is at the level of $1.8\sigma$ after accounting for the look-elsewhere effect \cite{Sohn:2024se}.
Although no statistically significant detection has yet been achieved, the pioneering studies have demonstrated the feasibility 
of testing cosmological colliders.
These developments motivate us to ask: 

\smallskip
\centerline{
    {\it What is the most significant collider signal in the latest dataset?}
}

\smallskip
At a more technical level, observational tests of the cosmological collider face two more challenges. 
On the one hand, the collider templates are complicated in general.
Their exact expressions,
normally in terms of special functions and infinite power series, are not well-suited for data analysis.
Furthermore, the large parameter space associated with particle masses, spins, and sound speeds poses significant challenges for both template construction and statistical interpretation.
For the CMB data analysis, more general bispectrum estimators, including the Modal estimator \cite{Fergusson:2008ra,Fergusson:2009nv, Fergusson:2010dm, Liguori:2010primordial, Fergusson:2014gea} and the modal KSW estimator \cite{Sohn:2023fte}, the generalised KSW approach \cite{Meerburg:2010osc,Munchmeyer2014kswosc}, binned estimators \cite{Bucher_2010,Philcox:2023binnedbs,Philcox:2023binnedbstensor}, and most recently a neural-network based KWS-type estimator \cite{Philcox:2025bbo} have enabled broader and less restrictive surveys of the bispectrum space. 
It remains to be seen which suits best for testing cosmological colliders.
On the other hand, as shown in previous studies \cite{Cabass:2024wob,Sohn:2024se}, there can be significant overlaps between the collider shapes and the equilateral-type non-Gaussianities that arise in single field inflation. 
Careful treatment of degeneracies with single-field shapes is essential to ensure that any apparent signal can be robustly attributed to genuine new physics effects.
Given the present observational status, an important question arises: 
\begin{center}
    {\it What is the best strategy for testing cosmological colliders at the current stage?}
\end{center}

In this work, targeting the questions above, we present an updated CMB analysis of cosmological collider signals that makes full use of the currently available \textit{Planck} data. Key ideas and results are highlighted in a companion paper \cite{Suman:2025vuf}.
Built on our previous work \cite{Sohn:2024se}, the present analysis incorporates several major improvements:
\begin{itemize}
    \item First, based on the minimal assumptions of weak couplings and scale invariance, we construct a more complete set of collider templates, spanning a wide range of particle masses, spins, and sound speeds. As the starting point of the bootstrap analysis is basic principles (unitarity, locality and spacetime symmetries), the result captures all the possible single-exchange bispectra with any interation, mass and spin.\footnote{To be more concrete, we limit our consideration to spin $\leq 2$, and the specific choices of interactions are presented in Section \ref{sec:scalar-osc} and Section \ref{sec:spin}. Other (higher-derivative) interactions can be generated by these examples through integration by parts.}
    \item Second, in order to isolate the distinctive signatures of massive particles and break the degeneracies with single-field PNG, we propose an {\it orthogonalization} procedure to reduce the overlaps between the collider shapes and two standard templates (both the equilateral and orthogonal ones). This leads to a full menu of orthogonalized collider templates. 
    \item Third, we perform the data analysis using the Modal pipeline, a principal bispectrum estimator employed and extensively validated for the \textit{Planck} analysis, and independent of our earlier analysis using the CMB-BEST pipeline. We also perform the cross-validation of the two estimators and provide consistency checks of the data analysis for certain collider signals.
\end{itemize}
Then final results are summarized in Table \ref{tab:constraint_summary}. 
We show that, in the Planck data, the most significant signal of $2.35\sigma$ comes from a massive-scalar exchange template,  after taking into account the look-elsewhere effects.
For aspects of methodology, our analysis establishes a systematic strategy optimized for probing cosmological colliders at the current stage of observations.
Thus for future observations, in addition to the detection of PNG, the orthogonalization approach proposed here may also provide the opportunity to rule out single field inflation,
as the new collider templates cannot be generated in single field models.

\begin{table}[ht]
    \centering
    \begin{tabular}{c c c c c c}
    \hline
    \hline
    \vspace{5pt}
         \textbf{Shape} & \textbf{Template} & \textbf{$f_{\rm NL}$} & \textbf{Raw} & \textbf{Adjusted} &
         \textbf{Best-Fit} \\
         [-0.3em]
          & &\textbf{ ($68\%$ CL)} & \textbf{SNR} & \textbf{SNR} & \textbf{Parameters} \\
    \hline
    \hline
    Scalar-I & \eqref{scalarI} & $182\pm 73$ & $2.48$ & $1.53$ & $\mu = 3.88$, $c_s = 45.82$ \\
    (\textit{orthogonalized}) &  & $3566\pm 1310$ & $2.72$ & $1.71$ & $\mu \lesssim 1.00$, $c_s = 0.12$ \\ 
     \hline
    Scalar-II & \eqref{scalarIIa} & $-295\pm 95$ & $3.09$ & $2.14$ & $\mu = 2.36$, $c_s = 0.12$ \\
   (\textit{orthogonalized}) &  & $-600\pm 182$ & $3.29$ & $2.35$ & $\mu = 1.85$, $c_s = 0.089$ \\ 
    \hline
    Spin-1 & \eqref{spin1m_simply} & $-132\pm 51$ & $2.58$ & $1.86$ & $\mu = 3.80$, $c_s = 2.76$ \\
    (\textit{orthogonalized}) &  & $483\pm 158$ & $3.05$ & $2.14$ & $\mu = 2.78$, $c_s = 1.73$  \\ 
    \hline
    Spin-2 & \eqref{spin2m} & $19\pm 7$ & $2.95$ & $1.93$ &  $\mu = 1.09$, $c_s = 0.68$\\
    (\textit{orthogonalized}) &  & $70\pm 22$ & $3.13$ & $1.98$ & $\mu = 2.10$, $c_s = 0.065$ \\
    \hline
    \end{tabular}
    \caption{Summary of CMB constraints on cosmological collider signals presented in this work. All templates depend on two parameters which were set to give the maximum absolute signal-to-noise ratio (raw). We also quote the signal-to-noise ratio adjusted for look elsewhere effect, explained in more detail in Section \ref{sec:look-elsewhere}. Each template quotes two constraints: the upper row corresponds to the theoretical shape as derived using bootstrap approach, while the bottom corresponds to a linear combination of the given template with equilateral and orthogonal shapes to generate an uncorrelated signal with these standard templates. This procedure is explained in Section \ref{sec:ortho}.}
    \label{tab:constraint_summary}
\end{table}

The paper is organized as follows. In Section \ref{sec:shapes} we first present a full menu of the cosmological collider bispectrum templates under the minimal assumptions of scale invariance and weak coupling.
Our consideration includes the three-point functions from single-exchange diagrams with any mass, spin, and interactions. In particular, in order to break the degeneracy with the single field EFT background signals,  we orthogonalize the shapes to be distinguishable from equilateral-type PNG.
In Section \ref{sec:cmbb}, we give an overview about the observational test of the CMB bispectrum statistics, with a focus on two data analysis pipelines: the Modal estimator and the CMB-BEST. Highlighting the Modal method in this work, we shall also provide a detailed comparison and consistency check with the previously CMB-BEST results on several benchmark examples.
In Section \ref{sec:planck}, we present the results of data analysis using the Modal pipeline and the Planck legacy release on the cosmological collider templates in Section \ref{sec:shapes}.
The summary and outlook are presented in Section \ref{sec:concl}.
We leave details of the statistical joint analysis about how to break degeneracies in Appendix \ref{app:joint}.

\section{Collider Templates}
\label{sec:shapes}

In this section, guided by the bootstrap computation, we propose a set of PNG templates for cosmological collider signals with various mass, spin and sound speed.
In particular, we try to design these templates to be simple but distinctive, such that they can be easily tested using observational data and reflect the true signatures of cosmological colliders.

\subsection{Progress and Problems}
The main observational target from inflation is the primordial curvature perturbation $\zeta$, which is related to the inflaton field fluctuation $\phi$ through $\zeta= (H/\dot\Phi)\phi$. 
Following the standard definition, its  bispectrum is introduced as
\begin{align}
\langle \zeta_{{\bf k}_1} \zeta_{{\bf k}_2} \zeta_{{\bf k}_3}  \rangle = (2\pi)^3 \delta({\bf k}_1+{\bf k}_2+{\bf k}_3) \frac{18}{5}\fnl \frac{S(k_1,k_2,k_3)}{k_1^2 k_2^2 k_3^2} P_\zeta^2~
,
\end{align}
with $ P_\zeta$ being the power spectrum of $\zeta$ and $\fnl$ the PNG size parameter. 
Our main focus is on the dimensionless shape function $S(k_1,k_2,k_3)$, which demonstrates rich $k$ dependence reflecting the various new physics effects during inflation.

On the theory side, one important goal is to identify all the possibilities of $S(k_1,k_2,k_3)$ from inflation. While this question may not have a concrete answer in the conventional model-dependent analysis, recently with the help of the cosmological bootstrap, significant progress has been made in deriving the most general form of the shape function. 
In this new approach, the analysis is based on the requirements of basic principles, which in the end leads to a systematic classification of cosmological correlators. 
For the scalar bispectrum, it turns out that two minimal assumptions are sufficiently constraining to find a complete set of shape functions. 
The first one is scale invariance, which eliminates feature models and requires the dimensionless shape functions to be in terms of momentum ratios.
The second assumption is weak coupling, which means that the leading contributions to the bispectrum come from Feynman diagrams with fewest vertices.
With these two assumptions, we are cornered into the following two broad scenarios of inflation. 
\begin{itemize}
    \item 
One is single field inflation, which corresponds to contact diagrams of inflaton self-interactions.  
There are two typical interactions $\dot\phi^3$ and $\dot\phi(\partial_i\phi)^2$ arising from the effective field theory (EFT) of single field inflation.
The resulting shape functions are peaked at the equilateral limit and their linear combinations lead to the equilateral and orthogonal shapes of PNG. 
To simplify the numerical computation in data analysis, we normally introduce the product-separable equilateral and orthogonal templates as
\bea
\label{equil} S^{\rm equil} & \equiv &
-\frac{k_1^2}{k_2k_3}-\frac{k_2^2}{k_1k_3}-\frac{k_3^2}{k_1k_2}-2+\(\frac{k_1}{k_2}+5 {\rm\ perms}\)\\
\label{ortho} S^{\rm ortho} & \equiv & 
-3\frac{k_1^2}{k_2k_3}-3\frac{k_2^2}{k_1k_3}-3\frac{k_3^2}{k_1k_2}-8+3\(\frac{k_1}{k_2}+5 {\rm\ perms}\)
\eea
If we consider higher-derivative interactions of the inflaton fluctuation, there will be infinite possible shapes from single field inflation, which have been systematically classified in the boostless bootstrap \cite{Pajer:2020wxk}. However, for the purpose of observational tests, these shapes generally have large overlaps with the equilateral and orthogonal templates. Thus within single field inflation, it is sufficient to take the general shape function as a linear combination $S= x S^{\rm equi} + y S^{\rm ortho}$, which was used in the data analysis of the Planck Collaboration \cite{Akrami:2019izv}. 
We call these shapes from single field inflation as equilateral-type non-Gaussianities.

\item Another scenario is the one with new degrees of freedom in addition to the inflaton fluctuations. These extra species, as intermediate states, can contribute to the inflaton bispectra through single exchange diagrams. The resulting shape functions $S_{\rm s.e.}$ carry characteristic fingerprints of these new degrees of freedom, such as their mass and spin. This is the so-called cosmological collider physics, which gives us a great opportunity to do particle spectroscopy via cosmic inflation. As each intermediate state generates a distinctive bispectrum shape, we expect many more possibilities of PNG in this scenario. It leads to certain challenges in the data analysis for the cosmological collider. In the rest of this section we shall investigate how to parametrize these bispectrum templates effectively.
\end{itemize}

The upshot of the above analysis is that, if the primordial Universe is described by an EFT of inflation with massive particles, then the complete scalar bispectrum is expected to be a linear combination of 
single-field templates and massive-exchange shapes with various masses, spins and interactions
\be \label{generalS}
S(k_1,k_2,k_3)= x S^{\rm equi} + y S^{\rm ortho} + \sum_n z_n S^{(n)}_{\rm s.e.}~,
\ee
where the sum is for all possible shapes from massive exchanges. The free parameters $x$, $y$ and $z_n$ are determined by the Wilson coefficients in the EFT.

Generally speaking, one may expect that the equilateral and orthogonal shapes act as background signals of PNG, while on the top of that we see the characteristic imprints of massive particles during inflation.
As in particle collider experiments, ideally we may first need to have a good measurement for the background part of the signal, and then move on to search for the presence of new heavy particles.
This can be done in principle if we have a sufficiently large amount of reproducible, high signal-to-noise data like at LHC.
However, constraining minuscule amounts of PNG imprinted nearly 14 billion years ago from observations of a single Universe realization is challenging, and we are unfortunately nowhere near that dream scenario for the detection of cosmological collider signals.
The latest CMB bispectrum analysis of the Planck experiment placed tight constraints on the size of the equilateral and orthogonal PNGs, but their amplitudes are statistically consistent with zero. There is no significant evidence for bispectrum shapes that can be seen as an equilateral-like background with small oscillations in the squeezed corner.

In the low signal-to-noise regime of PNG analysis, one cannot hope to study all possible terms $S^{(n)}_\mathrm{s.e.}$ and constrain $z_n$'s at once. Instead, characteristic signatures from cosmological colliders can be extracted to construct a template and fitted to the observational data, as shown in recent studies \cite{Cabass:2024wob,Sohn:2024se}.
In \cite{Sohn:2024se}, the cosmological collider templates have been constrained independently of the equilateral and orthogonal templates. While this approach is suitable for testing the full shape computed from a given physical scenario, the shape can also have sizeable contributions from the EFT shapes without collider-like oscillations. In principle, a large signal of the EFT shape can still result in a noticeable signal for the cosmological collider template.

The overlap in two templates can be quantified through their correlations (or cosine), defined through the following inner product with weight $w$:
\begin{align}
{ {\rm cos}\left( S^{(1)},S^{(2)} \right) } &\equiv \frac{\langle S^{(1)},  S^{(2)}\rangle}{\sqrt{\langle S^{(1)},  S^{(1)}\rangle \langle S^{(2)},  S^{(2)}\rangle}}, \quad {\rm where} \label{cosine} \\
\langle S^{(1)},  S^{(2)}\rangle &\equiv \iiint_{\mathcal{V}_k} w(k_1,k_2,k_3) S^{(1)}(k_1,k_2,k_3) S^{(2)} (k_1,k_2,k_3)  \mathrm{d}\mathcal{V}_k~,~~{\rm with}~ w\equiv \frac{1}{k_1+k_2+k_3}. 
\end{align}
This is a measure of similarity between the two shapes, with $1$ and $-1$ corresponding to perfect correlation (anti-correlation). Fitting two highly correlated shapes to the same observational dataset yields highly correlated $\fNL$ estimates. Their signal-to-noise ratios for $\fNL$ tend to be similar, and the two models cannot be distinguished easily from the data.
For cosmological collider, as we shall see below, many bispectrum shapes from massive exchanges are still mainly dominated by contributions that resemble the two single-field templates, and thus the cosine remains large. Thus, to truly search for the characteristic signatures of massive particles while being agnostic to the details of the single-field's effective action, it is important to account for this degeneracy with the equilateral-type PNGs.

With these subtleties in mind, 
we would like to identify the most efficient strategy for the observational test of cosmological colliders at the current stage.
More detailed discussions on possible approaches are left in Section \ref{sec:ortho}.

\subsection{Oscillations in the Squeezed Limit}\label{sec:scalar-osc}

Now we first take a closer look at the primordial bispectrum from massive scalar exchanges. For arbitrary quadratic and cubic interactions, the bootstrap computations of their exact shape functions are presented in Ref. \cite{Pimentel:2022fsc}, and two specific examples with simplified approximations are used in \cite{Sohn:2024se}. We shall not repeat the computation here, but highlight the generic features that can be useful for the data analysis.
Our main focus is on massive particles in the principal series, which usually lead to richer phenomenology in the PNG.

In general, we can separate the bispectrum shape from one particular single-exchange process into two different components: the EFT background and the collider signal
\be \label{sepa}
S_{\rm s.e.} = S_{\rm bkg} + S_{\rm col.}~.
\ee
This separation, differing from the one shown in the bootstrap computation,\footnote{There the full bispectrum can be expressed as a combination of homogeneous and inhomogeneous solutions of the boundary differential equations. The homogeneous solution $S_{\rm ho} \propto u^{1/2\pm i\mu}{}_2F_1$ contains the collider signal in the squeezed limit, where ${}_2F_1$s are hypergeometric functions that go to unity for $u\equiv k_3/(k_1+k_2)\rightarrow 0$. The inhomogeneous solution $S_{\rm inh}$ is an infinite power series that resembles the background component $S_{\rm bkg}$. Both of these two solutions contain unphysical singularities at $u\rightarrow 1$. To be more precise, we may link them to the two components in \eqref{sepa} as $S_{\rm col.}\propto u^{1/2\pm i\mu}$ and $S_{\rm bkg}=S_{\rm inh}+S_{\rm ho}-S_{\rm col.}$.} is based on their general properties and physical meanings, which are summarized as follows.
\begin{itemize}
    \item 
The $S_{\rm bkg}$ part can be approximately taken as rational polynomials of momentum variables, which resemble the bispectra from single field inflation, such as the equilateral and orthogonal shapes. Actually, when the intermediate states are much heavier than the Hubble scale of inflation, we find this part reduces to the single-field EFT predictions, which simply means that the heavy fields can be effectively integrated out in the low-energy theory. 
When the intermediate scalar is not so heavy, this $S_{\rm bkg}$ part becomes certain deformations of the single-field EFT shapes that are controlled by the mass parameter. 
One general behavior of $S_{\rm bkg}$ is the equilateral-type scaling in the squeezed limit $\lim_{k_3\rightarrow0}S_{\rm bkg} \propto k_3/(k_1+k_2)$.

\item
The $S_{\rm col.}$ part corresponds to the truly unique signature of massive particle production during inflation, which cannot be mimicked by other physical effects. Schematically, it has the following non-analytic form
\bea
S_{\rm col.} &\propto & \(\frac{k_3}{k_1+k_2}\)^{\frac{1}{2}+i\mu} +c.c. + {\rm perms.}\\ 
&=& \(\frac{k_3}{k_1+k_2}\)^{\frac{1}{2}} \cos\[\mu \log\(\frac{k_3}{k_1+k_2}\) +\delta \] + {\rm perms.}~,
\eea
where $\mu$ is determined by the mass of the scalar particle $\mu\equiv \sqrt{m^2/H^2 - 9/4}$.
When this intermediate particle is heavy, i.e. $\mu\gg 1$, the size of the collider signal is usually suppressed by a Boltzmann factor $e^{-\pi\mu}$. Thus we will mainly focus on cases with masses $\mu\sim\mathcal{O}(1)$.
Compared with the squeezed-limit scaling of $S_{\rm bkg}$, the collider signals are more dominant for $k_3\ll k_1\simeq k_2$, and thus they manifest themselves as oscillations in the squeezed limit of the bispectrum. 
\end{itemize}
 
One advantage of this separation is that we may absorb the $S_{\rm bkg}$ part into the equilateral-type shapes of single field non-Gaussianities, such that the distinctive collider signals from massive exchanges become manifest.
By doing so, the general form in \eqref{generalS} can be rewritten as
\be
S(k_1,k_2,k_3)= \tilde{x} S^{\rm equi} + \tilde{y} S^{\rm ortho} + \sum_n z_n S^{(n)}_{\rm col.}~.
\ee
Thus we use the linear combination of the equilateral and orthogonal templates to also mimic the background component from massive exchanges, and the sum is for the characteristic collider signals only. Next, we present two typical examples of $S_{\rm col.}$.

The first one is generated from an exchange diagram with the quadratic interaction $\dot\phi\s$ and cubic interaction $\dot\phi^2\s$. To have a general result, assume that the inflaton fluctuation has a sound speed $c_s$, and the massive field's sound speed is $c_\s$.
Then due to scale invariance, these two sound speeds would appear in terms of the ratio $c_s/c_\s$. In the following we simply use $c_s$ as this ratio, which can take any positive value.
As shown around eq.(5.16) in \cite{Pimentel:2022fsc}, the oscillatory collider signal from $\dot\phi^2\sigma$ is derived by acting the weight-shifting operator $\mathcal{W}_{12}=-c_s^2k_1k_2\partial^2_{k_{12}}$ on the non-analytic piece of the scalar seed function
\be
\langle \zeta_{{\bf k}_1} \zeta_{{\bf k}_2} \zeta_{{\bf k}_3}  \rangle 
\sim \frac{1}{k_1^2k_2^2k_3^2} \mathcal{W}_{12} \left[ \hat{\mathcal{I}}^{(0)}\left( \frac{k_3}{c_s(k_1+k_2)} \right) \right]+{\rm perm.} 
\ee
We shall not repeat the full steps here, but refer to (2.8)--(2.15) in \cite{Sohn:2024se} for more detailed derivation of this example.
The final result of the shape function is given by
\be \label{scalarI}
S_{\rm col.}^{\rm I}  = \frac{k_1k_2}{(k_1+k_2)^2} 
\(\frac{k_3}{k_1+k_2}\)^{1/2} \cos\[\mu\log\(\frac{k_3}{2c_s(k_1+k_2)}\) + \delta \] + 2~{\rm perms.}~, 
\ee     \label{eqn:template_SELM}
with the phase parameter
\be
\delta = \arg \[ \Gamma\(\frac{5}{2}+i\mu\)\Gamma\(-i\mu\) (1+i\sinh\pi\mu) \]~. \nn
\ee
We have stripped out the Boltzmann factor in \eqref{scalarI}, which contains the exponential suppression of heavy masses, and absorbed it into the normalisation of the shape function and $\fNL$.
While the conventional normalisation sets $S(k_*,k_*,k_*)=1$ at a pivot scale $k_*=0.05 \mathrm{Mpc}^{-1}$, the sinusoidal term in \eqref{eqn:template_SELM} can cause the shape function to vanish at this reference point. We therefore choose to normalise this template while setting $\cos=1$.
In the end, we have two free parameters in this template: the mass $\mu$ and the sound speed ratio $c_s$.
It is worth noting that
this shape function is similar with the {\it equilateral collider} template proposed in \cite{Pimentel:2022fsc}.
\begin{figure}
    \centering
    \begin{subfigure}[b]{1\textwidth}
        \centering
        \includegraphics[width=\textwidth]{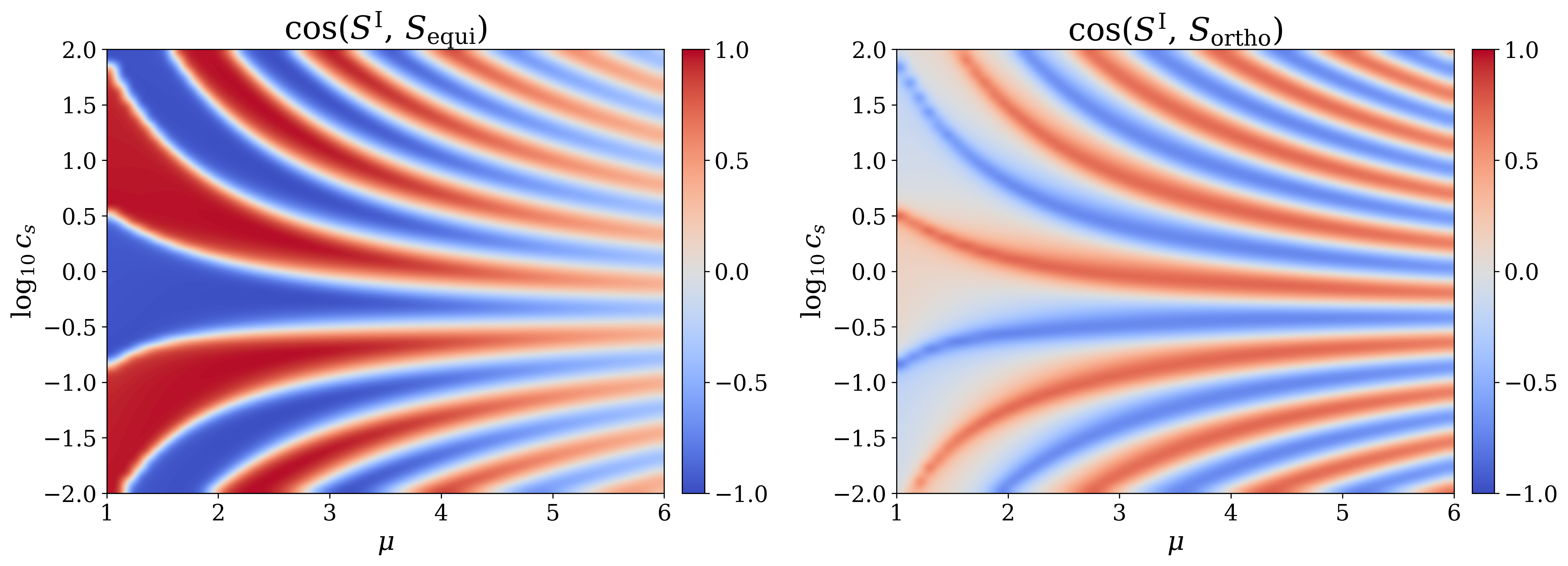}
    \end{subfigure}
    \hfill
    \begin{subfigure}[b]{1\textwidth}
        \centering
        \includegraphics[width=\textwidth]{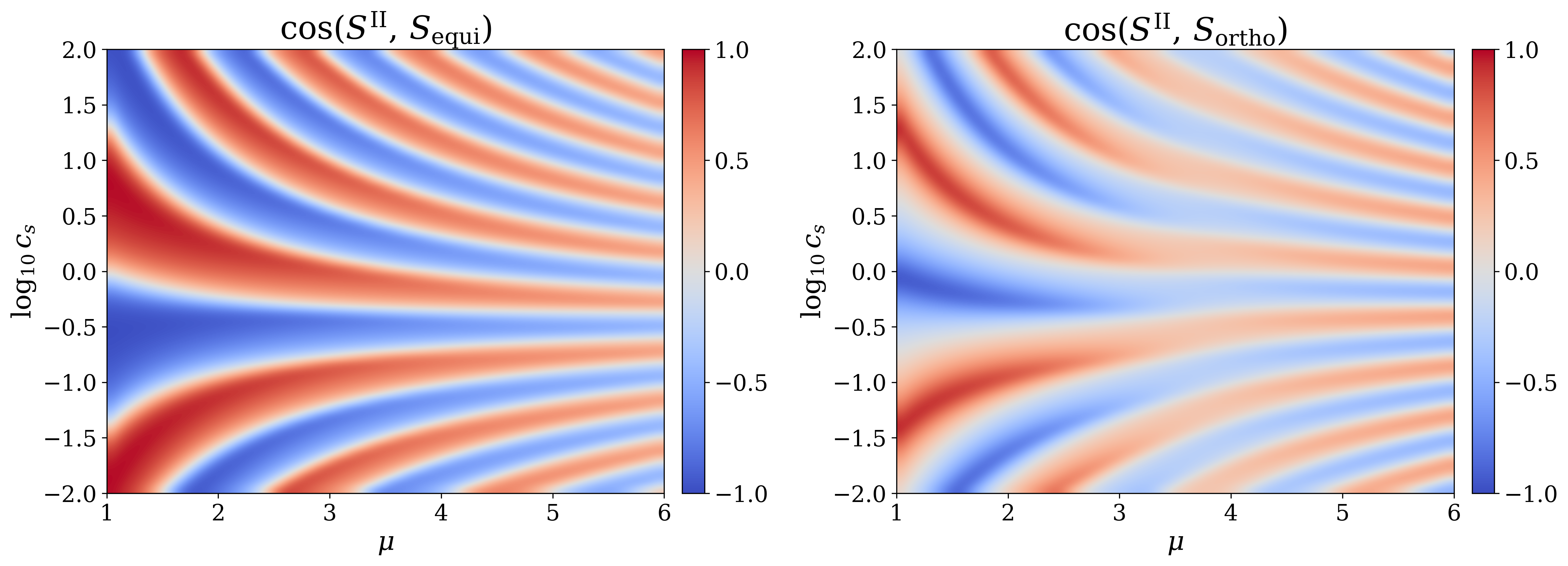}
    \end{subfigure}
    
    \caption{Shape cosine (correlation) in $k$-space between massive scalars and standard single-field shapes: equilateral \eqref{equil} and orthogonal \eqref{ortho} template.
    Top row depicts scalar-I \eqref{scalarI} correlations, and the scalar-II \eqref{scalarIIa} results are shown in the bottom row. All plots have the same color bar scale to accentuate the fact that massive scalar signals are overall less correlated with the orthogonal shape than with the equilateral.}
    \label{fig:corr_scalar}
\end{figure}

The second scalar-exchange template comes from the same $\dot\phi\s$ linear mixing but a different cubic vertex $(\partial_i\phi)^2\sigma$. 
Likewise, the oscillatory collider signal is given by
\bea \label{scalarIIa}
S_{\rm col.}^{\rm II} &=& \frac{k_3^2-k_1^2-k_2^2}{k_1k_2} \(\frac{k_3}{k_1+k_2}\)^{1/2} \left\{ \cos\[\mu\log\(\frac{k_3}{2c_s(k_1+k_2)}\) + \delta_1 \] \right.\nn\\
&& \left. + \sqrt{\beta} \frac{k_1k_2}{(k_1+k_2)^2} \cos\[\mu\log\(\frac{k_3}{2c_s(k_1+k_2)}\) + \delta_2 \]  \right\}    + 2~{\rm perms.}~, 
\eea    
with $\beta \equiv \mu^2+1/4$ and two phase parameters determined by the $\s$ mass
\bea
\delta_1 &=& \arg \[ \(\Gamma\(\frac{1}{2}+i\mu\) + \Gamma\(\frac{3}{2}+i\mu\)\)\Gamma\(-i\mu\) \(i+{\frac{1}{\sinh\pi\mu}}\) \]~,\\
\delta_2 &=& \arg \[ \Gamma\(\frac{5}{2}+i\mu\) \Gamma\(-i\mu\) \(i+{\frac{1}{\sinh\pi\mu}}\) \]~.
\eea
Again, we have neglected the overall prefactor with Boltzmann suppression, and normalize the shape function in the following way: first set two cosines to be unity, and then use the equilateral limit for normalization.

These two templates are expected to provide a basis for collider signals from massive scalar exchanges, as the role of templates \eqref{equil} and \eqref{ortho} for single field PNG.
It is important to notice that, although we have separated out the background components, the collider signals are not necessarily orthogonal to the equilateral-type shapes.
We can check how large the overlaps are between the two scalar-exchange collider shapes and the equilateral and orthogonal templates in \eqref{equil} and \eqref{ortho}. The shape cosines are shown in Figure \ref{fig:corr_scalar} as functions of the two free parameters $c_s$ and $\mu$. We see that for certain parameter choices (especially the low mass region) these two collider templates still resemble the single field PNG.

\subsection{Angular Dependence from Spinning Fields}\label{sec:spin}

For spinning exchanges, another typical signal of cosmological collider arises in the primordial bispectrum, which corresponds to the angular dependence in terms of Legendre polynomials.
For a massive spin-$s$ field $\s_{i_1...i_s}$, its linear mixing with the inflaton fluctuation requires that the only nonzero contribution to the scalar bispectrum comes from its longitudinal mode with helicity-$0$. 
As a result, the contraction between the helicity-$0$ polarization tensor $\epsilon^L_{i_1...i_s}$ and the external momentum leads to the following profile for one permutation of the scalar bispectrum 
\be
k_2^{i_1}k_2^{i_2}...k_2^{i_s}\epsilon^L_{i_1i_2...i_s}({\bf k}_3) \sim P_s(\hat{\bf k}_2\cdot \hat{\bf k}_3)~,
\ee
where the Legendre polynomial is a function of the angular between ${\bf k}_2$ and ${\bf k}_3$ and its index tells us about the spin of the exchanged intermediate state.
The full bispectrum shapes for spinning exchanges are in general quite complicated. Especially for higher spins, the computation involves spin-raising operators with multiple derivatives.
In the following, we shall take massive spin-$1$ and spin-$2$ particles as specific examples, and propose simplified templates that can reflect their characteristic collider signal.

Like massive scalar exchanges, the spinning-exchange bispectra can also be separated into two components: the non-oscillating background and the collider signal. The background part remains similar to the equilateral-type shapes of single field inflation in spite of the angular dependence \cite{MoradinezhadDizgah:2018ssw,Sohn:2024se}. Thus, the collider signal, which contains both squeezed-limit oscillations and angular-dependent profiles, are the true signatures of spinning particles.
For heavy fields with $\mu\gg1$, the signal is again exponentially suppressed by the Boltzmann factor. Thus we still focus on fields with masses of $\mathcal{O}(H)$.
We also take into account the sound speed parameters for both the inflaton and $\s_{i_1...i_s}$, and use their ratio in the templates.
For normalization, we follow the previous approach by first setting the oscillating cosines to be unity and then normalizing the templates at the equilateral configuration.
 
\begin{figure}
    \centering
    \includegraphics[width=1\linewidth]{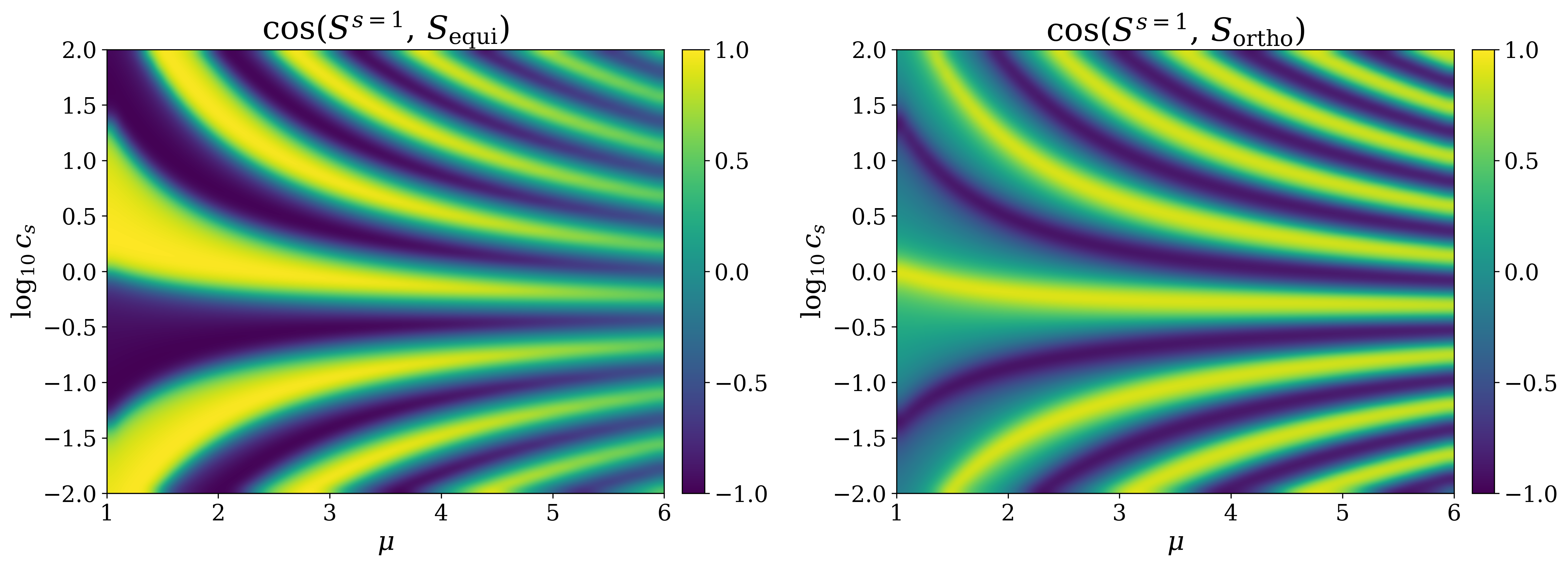}
    \caption{Shape cosine between the massive spin-1 \eqref{spin1m_simply} and standard equilateral and orthogonal templates. One can note a weaker correlation with the latter, a feature common with the scalar field templates.}
    \label{fig:corr_spin-1}
\end{figure}

With the above consideration, we apply the bootstrap computation for the full bispectrum shape from spinning exchanges. We focus on the quadratic and cubic interactions with lowest derivatives, which are expected to be the leading contributions in the EFT of cosmological collider \cite{Lee:2016vti}. 
For massive spin-1 particle $\s_i$, the couplings take the following form: $\partial_i\phi\s_i$ and $\dot\phi\partial_i\phi\s_i$.
We extract the following collider template from the full bootstrap computation \cite{Pimentel:2022fsc}
\bea \label{spin1m_simply}
S_{\rm col.}^{s=1} &\propto & 
P_1(\hat{\bf k}_2\cdot\hat{\bf k}_3)\frac{k_1}{(k_1+k_2)^2}  \(\frac{k_3}{k_1+k_2}\)^{1/2}  
 \left\{ k_1\cos\[\mu\log\(\frac{k_3}{2c_s(k_1+k_2)}\) 
+ \delta_1^{s=1} \] \right. \nn \\
&~& ~~~~~~~~~~~~~~~~~~~~~~\left.
 +  \sqrt{\beta+6} k_2 \cos\[\mu\log\(\frac{k_3}{2c_s(k_1+k_2)}\) + \delta_2^{s=1}  \]  \right\}    +    5~{\rm perms.}~, 
\eea    
with two phase parameters
\bea
\delta_1^{s=1} &=& \arg \[ {\Gamma\(-i\mu\)}{\Gamma\(\frac{1}{2}+i\mu\)} \(1+{i\sinh\pi\mu}\) \]~,\\
\delta_2^{s=1} &=& \arg \[ {\Gamma\(-i\mu\)}{\Gamma\(\frac{1}{2}+i\mu\)} \(1+{i\sinh\pi\mu}\) \(\frac{5}{2}+i\mu\)\]~.
\eea
In Figure \ref{fig:corr_spin-1}, we show the correlations with the equilateral and orthogonal shapes by varying the two free parameters $\mu$ and $c_s$.

\begin{figure}
    \centering
\includegraphics[width=1\linewidth]{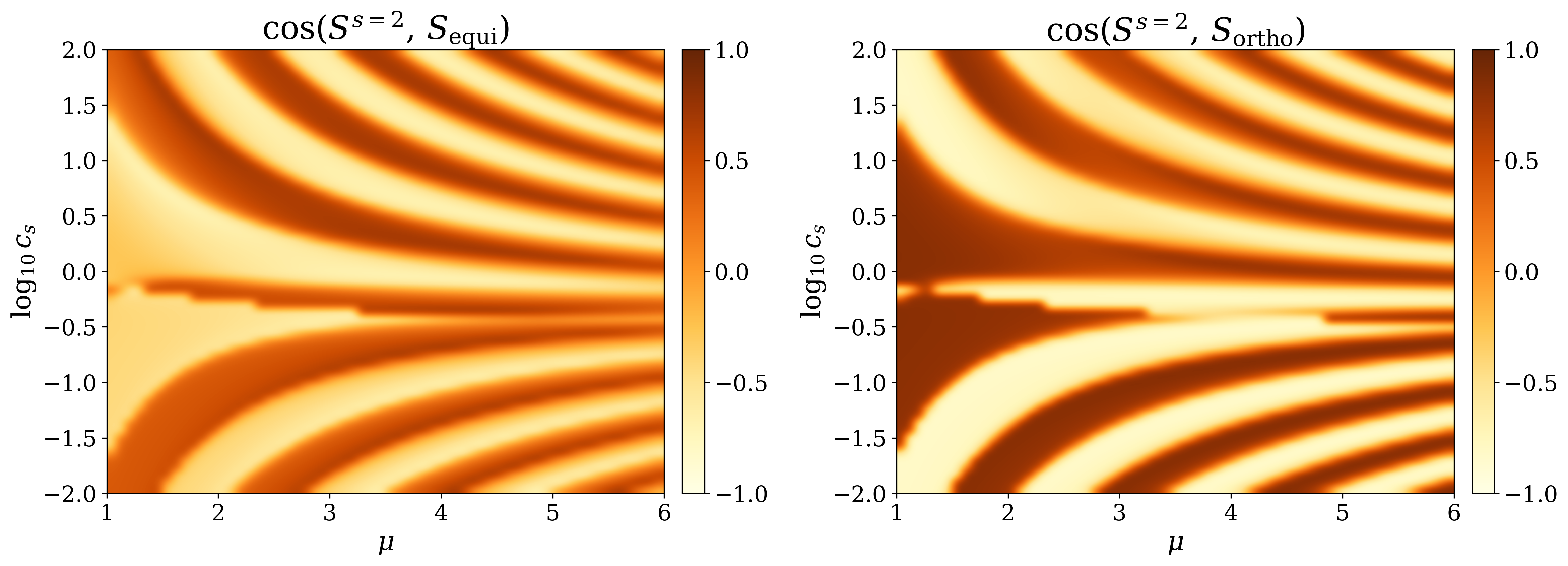}
    \caption{Shape cosine between the massive spin-2 \eqref{spin2m} and standard equilateral and orthogonal templates, with the former being less correlated as opposed to other templates. Apparent `glitch' and asymmetry with respect to the $\log c_s$ stems from the fact that the spin-2 template contains two terms of different powers of $c_s$ which cannot be removed by normalization.}
    \label{fig:corr_spin-2}
\end{figure}

Similarly for massive spin-2 particle $\s_{ij}$, the leading EFT interactions are $\partial_{ij}\phi\s_{ij}$ and $\dot\phi\partial_{ij}\phi\s_{ij}$.
The collider template is extracted from the bootstrap computation in \cite{Pimentel:2022fsc}
\begin{small}\bea \label{spin2m}
S_{\rm col.}^{s=2} &=& P_2(\hat{\bf k}_2\cdot\hat{\bf k}_3)\frac{k_1k_2}{8(k_1+k_2)^3}  \sqrt{\frac{\pi^3\[1+\sinh^2(\pi\mu)\]}{2\mu \tanh(\pi \mu) }} \frac{\beta^2(\beta+2)^2}{\cosh^2(\pi\mu)} \(\frac{k_3}{c_s(k_1+k_2)}\)^{1/2} \nn \\
&&\times \left\{ k_1\cos\[\mu\log\(\frac{k_3}{2c_s(k_1+k_2)}\) + \delta_1^{s=2} \]
 +  \sqrt{\beta+12} k_2 \cos\[\mu\log\(\frac{k_3}{2c_s(k_1+k_2)}\) + \delta_2^{s=2}  \]  \right\}   \nn
\\ 
&& -P_2(\hat{\bf k}_2\cdot\hat{\bf k}_3)\frac{k_1k_2}{6(k_1+k_2)^3}  \sqrt{\frac{\pi^3\beta(\beta+2)(\beta+6)(\beta+12)}{\mu \tanh(\pi \mu) }} \frac{\beta(\beta+2)}{\cosh(\pi\mu)} \(\frac{k_3}{2c_s(k_1+k_2)}\)^{5/2} \nn \\
&&\times \left\{ k_1\cos\[\mu\log\(\frac{k_3}{2c_s(k_1+k_2)}\) + \delta_3^{s=2} \]
 +  \sqrt{\beta+30} k_2 \cos\[\mu\log\(\frac{k_3}{2c_s(k_1+k_2)}\) + \delta_4^{s=2}  \]  \right\} \nn \\
 && +     5~{\rm perms.}~, 
\eea    \end{small}
with 
\bea
\delta_1^{s=2} &=& \arg \[ \frac{\Gamma\(-i\mu\)}{\Gamma\(\frac{1}{2}-i\mu\)} \(1+{i\sinh\pi\mu}\) \]~,\\
\delta_2^{s=2} &=& \arg \[ \frac{\Gamma\(-i\mu\)}{\Gamma\(\frac{1}{2}-i\mu\)} \(1+{i\sinh\pi\mu}\) (7+2i\mu)\]~,\\
\delta_3^{s=2} &=& \arg \[  {\Gamma\(-i\mu\)}{\Gamma\(\frac{9}{2}+i\mu\)} \(1+{i\sinh\pi\mu}\) \]~,\\
\delta_4^{s=2} &=& \arg \[ {\Gamma\(-i\mu\)}{\Gamma\(\frac{9}{2}+i\mu\)} \(1+{i\sinh\pi\mu}\) \(\frac{11}{2}+i\mu\)\]~.
\eea
Its overlap with the equilateral and orthogonal templates is shown in Figure \ref{fig:corr_spin-2} for certain parameter ranges of $\mu$ and $c_s$.

\subsection{Orthogonalizing Collider Templates}
\label{sec:ortho}

Based on the EFT consideration, we expect that the sum of collider signals in the full primordial bispectrum \eqref{generalS} should include the above four templates \eqref{scalarI}, \eqref{scalarIIa}, \eqref{spin1m_simply} and \eqref{spin2m} as leading components.
Meanwhile, as we have seen in Figure \ref{fig:corr_scalar}, \ref{fig:corr_spin-1} and \ref{fig:corr_spin-2}, there can still be large overlaps between some collider templates and the equilateral-type non-Gaussianities. 
Since these bispectrum templates include \textit{both} the signatures of massive particles and some EFT background, the corresponding observational tests also depend on the single-field predictions.

The collider signals can be constrained independently from the EFT background using two approaches: marginalization and orthogonalization. In the following, we demonstrate the methods using a simpler example. A more formal introduction to statistics of joint bispectrum constraints is detailed in Appendix \ref{app:joint}.

\begin{figure}
  \centering
\includegraphics{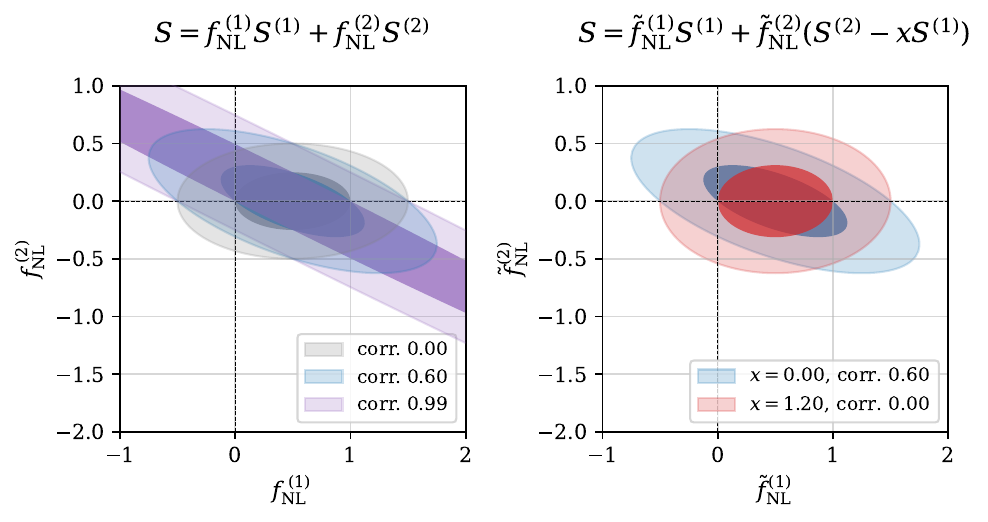}
    \caption{Illustrations of $1\sigma$ and $2\sigma$ contour plots for joint $\fnl$ constraints of two bispectrum shapes. \textit{Left}: joint constraints on two shapes are shown with varying amounts of correlations between them: 0 (grey), 0.6 (blue), and 0.99 (purple). Note that a positive correlation between the shapes induces an anticorrelation
between the estimated $\fnl$s, which is why the contours tilt. Independent analysis of $\fnl^{(2)}$ corresponds to setting $\fnl^{(1)}=0$, and is equivalent to the marginalized constraints if the two shapes are uncorrelated (grey contours). \textit{Right}: joint constraints for a fixed pair of shapes, where the second shape now includes contributions from the first through the redefinition
$\tilde{S}^{(2)} = S^{(2)} - x S^{(1)}$, shown for $x=0$ (blue, identical to the left panel) and
$x=1.2$ (red, the orthogonalized combination). Appropriate choice of $x$ effectively orthogonalizes the shape (red contours), making it uncorrelated with the first shape.
    }
    \label{fig:ellipse}
\end{figure}

\begin{framed}
\noindent 
\paragraph{Marginalization {\it vs} Orthogonalization:}{\it a two-template example}.

\vskip6pt
Here, we review the statistical formalism regarding bispectrum analysis involving multiple templates, using a simplified example. Suppose that the bispectrum is a linear combination of two templates:
\be \label{2shapes}
S=\fnl^{(1)}S^{(1)} + \fnl^{(2)}S^{(2)}~.
\ee
Here, we assume that the observational constraints provide a multivariate normal distribution for the two $\fnl$s, which is shown by the contours in the left panel of Figure \ref{fig:ellipse}.
The three cases shown in the plot have identical constraints $\fnl^{(1)} = 0.5\pm0.5$ and $\fnl^{(2)} = 0\pm0.25$ at the $1\sigma$ level (68\% confidence level, CL hereafter) for the two $\fnl$s individually, but have varying levels of correlation between the two shapes.

\textit{Joint} constraints measure how likely it is to have $(\fnl^{(1)},\fnl^{(2)})$ given the data. The constraints are represented by the multivariate normal likelihood, which can be visualized as inner and outer ellipses as shown in the plot, corresponding to the $1\sigma$ (68\% CL) and $2\sigma$ (95\% CL) contours, respectively. Note that positive shape correlations result in anti-correlations between the $\fNL$ estimates. 

\textit{Marginalized} constraints represent the probability of measuring $\fnl^{(2)}$ given the data when no prior knowledge on $\fnl^{(1)}$ is assumed. These can be obtained by integrating the likelihood by $\fnl^{(1)}$ with a flat prior, effectively projecting the contours in the plot onto the $y$ axis. 

\textit{Conditional} constraints on $\fnl^{(2)}$ is evaluated under the condition $\fnl^{(1)}=c$ for some $c$. The probability distribution can be evaluated along the line $\fnl^{(1)}=c$ with a suitable normalization factor. The case $c=0$ yields \textit{independent} constraints where the model is $S=\fnl^{(2)}S^{(2)}$. Independent constraints come from single-template analysis and can also be interpreted as the constraint conditional on all other $\fnl$s vanishing.

When the two shapes are orthogonal, the $\fnl$s are uncorrelated like the grey ellipses in the plot. In this case, all conditional and marginalized constraints are identical. The joint $\fnl$ constraints are then equivalent to combining two independent constraints from fitting $S=\fnl^{(1)}S^{(1)}$ and $S=\fnl^{(2)}S^{(2)}$ separately.

When the shapes are not orthogonal, however, the corresponding $\fnl$ estimates are correlated. The joint constraints are represented by tilted elliptical contours (blue in the plot) or an infinite band for near-perfect correlation (purple). Note that a positive correlation between the shapes means a negative correlation of estimated $\fnl$s, as a larger $\fnl^{(1)}$ can be partially compensated by a smaller $\fnl^{(2)}$ and vice versa.

One way to constrain $\fnl^{(2)}$ while accounting for the partial degeneracy is to \textit{marginalize} over $\fnl^{(1)}$. However, marginalized constraints are no longer equivalent to conditional/independent constraints; not only do the central (best-fit) values of $\fnl^{(2)}$ differ, but their error bars are larger than those of the conditional constraints. This is because a larger part of the shape $S^{(2)}$ is degenerate with $S^{(1)}$, and the lack of information on $\fnl^{(1)}$ increases the uncertainty of the $\fnl^{(2)}$ estimation. For the extreme case where the two shapes are perfectly correlated, we have $S^{(1)}=rS^{(2)}$ for some constant $r$. The model can be written as $S=(r\fnl^{(1)} + \fnl^{(2)})S^{(2)}$ in this case, and the observational constraints can only place bounds on the particular combination $r\fnl^{(1)}+\fnl^{(2)}$. 

An alternative approach to studying the signatures of $S^{(2)}$ is to \textit{orthogonalize} the template: $\tilde{S}^{(2)}=S^{(2)}-xS^{(1)}$ for some $x$ such that $S^{(1)}$ and $\tilde{S}^{(2)}$ are now orthogonal. This is depicted in the right panel of Figure \ref{fig:ellipse}. Since the $\fnl$ estimate of the orthogonalized shape is now uncorrelated with $\fnl^{(1)}$, independent constraints on $\fnl^{(1)}$ and $\tilde{f}^{(2)}_\mathrm{NL}$ convey the full information from the joint constraints, and are equivalent to the marginalized constraints.
\end{framed}

We would like to search for signatures of the cosmological collider templates $S^\mathrm{col.}$ with some unknown single-field contributions $S^\mathrm{EFT}$ which has notable overlap with $S^\mathrm{col.}$. Both strategies above\textemdash marginalization and orthogonalization\textemdash can help us account for the partial degeneracy and identify the characteristic signatures of massive particles during inflation.

In \cite{Cabass:2024wob}, the spin-0 cosmological colliders signatures were studied jointly with the EFT background with BOSS galaxy power spectrum and bispectrum, where the $\fnl^\mathrm{equil}$, $\fnl^\mathrm{equil}$, and other parameters related to the likelihood have been marginalized over. \footnote{This work also included some \textit{Planck} CMB bispectrum constraints, assuming that the collider template is completely represented by a linear combination of the equilateral and orthogonal templates. This assumption ignores all characteristic signatures of the cosmological collider predictions \textit{not} captured by single-field EFT, the main target of our work.}

In this work, we will first orthogonalize each collider template with the single-field shapes and then constrain them using the \textit{Planck} CMB bispectrum. Note that this is analogous to how the standard `orthogonal' template was first introduced to be orthogonal to the local and equilateral templates \cite{Senatore:2009gt}. There are several reasons for this choice:

\begin{itemize}
    \item \textit{Reproducibility}. Since the template has a fixed form, a) the theoretical predictions of the primordial bispectum shape can be directly compared with the observational constraints, and b) constraints from different observational tests can be compared with ease. 
    \item \textit{Low signal-to-noise regime}. All observational data to date are consistent with zero PNG. We focus on searching for the signatures of collider signals ($\fnl\neq0$) for various possible scenarios using uncorrelated templates. The remaining correlations between different parameter choices are then accounted for through look-elsewhere effect analysis, as detailed in Section \ref{sec:look-elsewhere}.
    \item \textit{Prior independent}. Marginalization requires a choice of prior, and while assuming a uniform prior for $\fnl$s can be a safe choice, a log-uniform prior may make more sense for $\fnl$s and theory parameters that relate to unknown coupling coefficients in the action. We stay agnostic to this choice by not marginalizing over and analyzing independent constraints directly.
\end{itemize}

The orthogonalization of collider templates proceeds as follows.
We seek to construct a special linear combination \eqref{sepa} which would give a shape function uncorrelated with both the equilateral and orthogonal templates 
\begin{equation} \label{ortho}
\begin{cases}
        \langle S_{\rm col.} + x S^{\rm equi} + y S^{\rm ortho}, S^{\rm equi} \rangle &=0 \\
        \langle S_{\rm col.} + x S^{\rm equi} + y S^{\rm ortho}, S^{\rm ortho} \rangle &=0,
\end{cases}
\end{equation}
where the inner product between shapes in $k$-space is defined in Equation \eqref{cosine}.

The requirement \eqref{ortho} gives us two equations such that we can fix the two coefficients $x$ and $y$ for each template.
Here we apply this orthogonalization procedure for the four collider templates \eqref{scalarI}, \eqref{scalarIIa}, \eqref{spin1m_simply} and \eqref{spin2m}. 
The coefficients $x$ and $y$ are computed numerically for each choice of parameters $(\mu, c_s)$ using the Modal decomposition \eqref{eqn:separable_basis_expansion}, as this is the format of bispectra that enters data analysis.

Note that the inner product in \eqref{ortho} is defined in the primordial $k$-space, which is independent of the observational data used. This inner product is, in general, different from the one for $\fnl$ estimates that is defined from Fisher information (see Appendix \ref{app:joint} for details). The $\fnl$ estimates from the orthogonalized templates are therefore not completely uncorrelated with $\fnl^\mathrm{equil}$ and $\fnl^\mathrm{ortho}$, unlike the right plot of Figure \ref{fig:ellipse}.

In Section \ref{sec:planck} we present \textit{Planck} CMB constraints on these orthogonalized templates, together with the original templates for comparison.

\section{CMB Bispectrum}\label{sec:cmbb}
Quantum fluctuations of the inflaton and any interacting fields during inflation are imprinted in the primordial curvature perturbation $\zeta$ which sources the metric perturbation on superhorizon scales. These primordial fluctuations re-enter the horizon after the end of inflation and affect the photon distribution at recombination, leaving observable imprints in the anisotropies of the cosmic microwave background (CMB). 

Although the two-point correlation function (power spectrum) of the CMB has been measured with percent level precision, providing stringent constraints on the parameters of the $\Lambda$CDM model, it is insensitive to certain classes of inflationary dynamics. In particular, signatures of nontrivial interactions and multi-field effects often manifest through non-Gaussian correlations. If the primordial fluctuations are non-Gaussian, the two-point correlation function, or alternatively the power spectrum, fails to fully describe the statistics of the CMB, and one needs to compute at least one order higher in the correlation functions to capture non-Gaussian features.

A nonzero primordial bispectrum, or the corresponding three-point correlation function, would offer compelling evidence for such inflationary physics beyond the simplest models. Despite recent advances in constraining PNGs using large-scale structure (LSS) surveys, CMB observations remain the most direct and robust probe due to their early-time origin, linear relation to primordial curvature perturbations, and minimal contamination from the late time universe.

In this section, we give an overview of CMB bispectrum statistics, specifically outlining differences between the Modal and CMB-BEST estimators. We explain the look-elsewhere-effect and how to translate the $f_{\rm NL}$ measurements into the language of particle physics detection. Finally, we revisit three distinct bispectrum templates with small sound speeds and cross-validate the two pipelines.

\subsection{CMB Bispectrum Statistics}
Inflationary models that predict PNGs are characterized by a shape function $S(k_1,k_2,k_3)$, which is related to the CMB bispectrum via the radiation transfer functions $T_\ell(k)$. The CMB bispectrum, defined as the three-point correlation function of temperature (and polarization) anisotropies $a_{\ell m}$, is given by
\begin{equation}
    B_{m_1,m_2,m_3}^{\ell_1,\ell_2,\ell_3} = \langle a_{\ell_1 m_1} a_{\ell_2 m_2} a_{\ell_3 m_3} \rangle = \mathcal{G}_{m_1,m_2,m_3}^{\ell_1,\ell_2,\ell_3} \, b_{\ell_1 \ell_2 \ell_3},
\end{equation}
where $\mathcal{G}_{m_1,m_2,m_3}^{\ell_1,\ell_2,\ell_3}$ is the Gaunt coefficient\footnote{$\mathcal{G}_{m_1,m_2,m_3}^{\ell_1,\ell_2,\ell_3} = \int \rm d^2\Omega Y_{\ell_1m_1}(\Omega)Y_{\ell_2m_2}(\Omega)Y_{\ell_3m_3}(\Omega)$} enforcing momentum conservation, and $b_{\ell_1 \ell_2 \ell_3}$ is the reduced bispectrum containing all the physical information about inflationary dynamics:
\begin{equation}\label{eq:reduced_bispectrum}
    b_{\ell_1 \ell_2 \ell_3} = \left( \dfrac{2}{\pi} \right)^3 \int \mathrm{d}r\, r^2 \iiint \mathrm{d}k_1 \mathrm{d}k_2 \mathrm{d}k_3 \, S(k_1,k_2,k_3) \prod_{i=1}^3 \left[ j_{\ell_i}(k_i r) \, T_{\ell_i}(k_i) \right].
\end{equation}
Computing the full bispectrum requires summing over three $\ell$-indices and two $m$-indices (with $m_3$ fixed by $m_1 + m_2 + m_3 = 0$), resulting in a computational complexity of $\mathcal{O}(\ell_{\rm max}^5)$. For high-resolution experiments such as \textit{Planck} ($\ell_{\rm max} \sim 2000$) and upcoming surveys like the Simons Observatory, this becomes computationally prohibitive. In addition, the signal-to-noise ratio for individual bispectrum components is typically too low to allow model-independent detection of individual multipole components.

As a result, observational analysis proceeds by taking a theoretical prediction from a specific inflation model
\begin{equation}\label{cmb_bispectrum_th}
    B_{m_1,m_2,m_3}^{\ell_1,\ell_2,\ell_3 (\rm th)} = f_{\rm NL} \, \mathcal{G}_{m_1 m_2 m_3}^{\ell_1 \ell_2 \ell_3} \, b_{\ell_1 \ell_2 \ell_3}^{(\rm th)},
\end{equation}
and comparing it directly against the observed bispectrum
\begin{equation}
    B_{m_1,m_2,m_3}^{\ell_1,\ell_2,\ell_3 (\rm obs)} = a_{\ell_1 m_1} a_{\ell_2 m_2} a_{\ell_3 m_3} - \left(\langle a_{\ell_1 m_1} a_{\ell_2 m_2}\rangle a_{\ell_3 m_3} + 2 \text{ perms}. \right),
\end{equation}
where the second term in the brackets accounts for off diagonal correlations caused by masking effects and anisotropic sky coverage, effectively acting as a mean field subtraction. The $f_{\rm NL}$ factor in \eqref{cmb_bispectrum_th} parameterizes the overall amplitude of a bispectrum shape. To put a constraint on a particular inflationary model using the CMB precisely means to estimate this $f_{\rm NL}$ value along with its corresponding error.

Hence, using a condensed notation\footnote{Short-hand notation following the literature: $\boldsymbol{l}_j  \coloneqq (l_j,m_j) ,\quad
    L \coloneqq (l_1,l_2,l_3) ,\quad
    \boldsymbol{L} \coloneqq (\boldsymbol{l_1},\boldsymbol{l_2},\boldsymbol{l_3})$}, the estimation problem reduces to a linear regression:
\begin{equation}
    B_{\boldsymbol{L}}^{(\mathrm{obs})} = B_{\boldsymbol{L}}^{(\rm th)} f_{\mathrm{NL}} + \epsilon_{\boldsymbol{L}}.
\end{equation}
The regression variables are further modified so that the resulting covariance matrix reduces to identity while keeping the estimator unbiased.
Assuming $a_{lm}$'s to be weakly non-Gaussian and noting that cosmic variance dominates over measurement errors, the variables become:
\begin{equation}
        \Tilde{B}_{\boldsymbol{L}}^{\mathrm{obs}} = \dfrac{a_{\boldsymbol{l_1}}a_{\boldsymbol{l_2}} a_{\boldsymbol{l_3}} - \langle a_{\boldsymbol{l_1}}a_{\boldsymbol{l_2}} \rangle a_{\boldsymbol{l_3}}
    - \langle a_{\boldsymbol{l_2}}a_{\boldsymbol{l_3}} \rangle a_{\boldsymbol{l_1}} -
    \langle a_{\boldsymbol{l_1}}a_{\boldsymbol{l_3}} \rangle a_{\boldsymbol{l_2}}}{\sqrt{\Delta_{\boldsymbol{l_1}\boldsymbol{l_2}\boldsymbol{l_3}}C_{l_1}C_{l_2}C_{l_3}}}
    \quad
    \text{and}
    \quad
    \Tilde{B}_{\boldsymbol{L}}^{(\rm th)} = \dfrac{\mathcal{G}_{\boldsymbol{L}}b_{L}^{(\rm th)}}{\sqrt{\Delta_{\boldsymbol{l_1}\boldsymbol{l_2}\boldsymbol{l_3}}C_{l_1}C_{l_2}C_{l_3}}},
\end{equation}
where $\Delta_{\boldsymbol{l_1}\boldsymbol{l_2}\boldsymbol{l_3}}$ is a symmetry factor either equal to 6 (if $\boldsymbol{l_1} =\boldsymbol{l_2} = \boldsymbol{l_3}$), equal to 2 (if two $\boldsymbol{l}$'s identical) or equal to 1 (if all three different). Theoretically, we would define $C_{\boldsymbol{l_1l_2}} = \langle a_{\boldsymbol{l_1}} a_{\boldsymbol{l_2}} \rangle$, but in practice we either take $\langle a_{\boldsymbol{l_1}} a_{\boldsymbol{l_2}} \rangle_{\textit{MC}}$ or approximate $C_{\boldsymbol{l_1l_2}} \approx C_l \delta_{\boldsymbol{l_1l_2}}$. With these variables, the error in each $\boldsymbol{L}$-mode is independent with unit variance. These variables now satisfy the necessary conditions to perform ordinary least squares regression (or equivalently, the maximum likelihood estimate) for $f_{\rm NL}$:
\begin{equation}\label{eq:fnl_estimator}
    \hat{f}_{\mathrm{NL}}
    =
    \dfrac{1}{F} \sum_{l_jm_j} \dfrac{\mathcal{G}^{l_1l_2l_3}_{m_1m_2m_3} b_{l_1l_2l_3}}{6C_{l_1}C_{l_2}C_{l_3}}  [a_{l_1m_1}a_{l_2m_2}a_{l_3m_3} - 3\langle a_{l_1m_1}a_{l_2m_2} \rangle a_{l_3m_3}]
    = \frac{P}{F},
\end{equation}
where $F$ and $P$ are given by
\begin{equation}
\begin{split}
    F &\equiv \tilde{B}^{\rm th^2}
    = 
    \sum_{\boldsymbol{l_1} \leq \boldsymbol{l_2} \leq \boldsymbol{l_3}} \dfrac{\mathcal{G}_{\boldsymbol{L}}^2 b_L^{(\rm th)^2}}{\Delta_{\boldsymbol{l_1}\boldsymbol{l_2}\boldsymbol{l_3}}C_{l_1}C_{l_2}C_{l_3}}
    =
    \sum_{\boldsymbol{l_1} \boldsymbol{l_2} \boldsymbol{l_3}} \dfrac{\mathcal{G}_{\boldsymbol{L}}^2 b_L^{(\rm th)^2}}{6C_{l_1}C_{l_2}C_{l_3}}
    = 
    \sum_{l_1 l_2 l_3} \dfrac{h_{L}^2 b_L^{(\rm th)^2}}{6C_{l_1}C_{l_2}C_{l_3}}\\
    P &\equiv \tilde{B}^{\rm th} \cdot \tilde{B}^{\mathrm{obs}} 
    = 
    \sum_{\boldsymbol{l_1} \boldsymbol{l_2} \boldsymbol{l_3}} \dfrac{\mathcal{G}_{\boldsymbol{L}} b_L^{(\rm th)}}{6C_{l_1}C_{l_2}C_{l_3}}
    [a_{\boldsymbol{l_1}}a_{\boldsymbol{l_2}} a_{\boldsymbol{l_3}} - \langle a_{\boldsymbol{l_1}}a_{\boldsymbol{l_2}} \rangle a_{\boldsymbol{l_3}}
    - \langle a_{\boldsymbol{l_2}}a_{\boldsymbol{l_3}} \rangle a_{\boldsymbol{l_1}} -
    \langle a_{\boldsymbol{l_1}}a_{\boldsymbol{l_3}} \rangle a_{\boldsymbol{l_2}}].
\end{split}
\end{equation}
The estimator \eqref{eq:fnl_estimator} is unbiased and optimal. If the underlying model is true, then $\langle \hat f_{\rm NL} \rangle = f_{\rm NL}$. Furthermore, the expected variance saturates the Cramer-Rao bound so is given by the Fisher error $\rm Var(\hat f_{\rm NL}) = \hat \sigma_{f_{\rm NL}}^2= F^{-1}$. The Fisher information naturally motivates the following definition of an inner product on the space of reduced bispectra:
\begin{equation}\label{eq:late-time-inner-product}
    \langle b^{(i)}, b^{(j)} \rangle
    \coloneqq
    \sum_L \dfrac{h_L^2 b_L^{(i)} b_L^{(j)}}{6C_{l_1}C_{l_2}C_{l_3}} = F_{ij}. 
\end{equation}
This enables a simple way to calculate the correlation between two CMB bispectrum shapes and their $f_{\rm NL}$ estimates
\begin{equation}\label{eqn:late_time_corr}
    \mathrm{Corr}\left( \hat f^{(i)}_{\rm NL}, \hat f^{(j)}_{\rm NL} \right) 
    =
    \mathrm{Corr}\left( b^{(i)}, b^{(j)} \right)
    =
    \dfrac{\langle b^{(i)}, b^{(j)} \rangle}{\sqrt{\langle b^{(i)}, b^{(i)} \rangle \langle b^{(j)}, b^{(j)} \rangle}}.
\end{equation}

\subsection{Modal CMB estimator}
Computing the CMB bispectrum from the primordial shape function in \eqref{eq:reduced_bispectrum}, requires evaluating a four-dimensional integral over the momenta $k_i$ and the radial variable $r$. In addition to the high computational cost scaling as $\mathcal{O}(\ell_{\mathrm{max}}^5)$, the numerical evaluation of this integral is challenging due to oscillatory integrands and the coupling of variables.

Significant simplification arises if the primordial bispectrum is separable in its momentum arguments $S(k_1,k_2,k_3) = X(k_1)Y(k_2)Z(k_3)$. In this case, the triple integral over $k_i$ reduces to a product of one-dimensional integrals. The remaining radial integral over spherical Bessel functions can then be performed efficiently. This separability is the key idea behind the fast Komatsu-Spergel-Wandelt (KSW) estimator \cite{Yadav:2007rk}.

Although the KSW approach is fast and optimal for separable templates, it is inherently limited to a restricted class of shapes that can be written in factorizable form. To overcome this limitation, the Modal estimator \cite{Fergusson:2009nv, Fergusson:2010dm, Liguori:2010hx, Fergusson:2014gea} generalizes the method by expanding arbitrary bispectrum shapes in a basis of separable functions
\begin{equation}\label{eqn:separable_basis_expansion}
    S(k_1,k_2,k_3) = \sum_{n=0}^{n_{\rm max}} \alpha_n  Q_n (k_1,k_2,k_3),
\end{equation}
where the $Q_n(k_1,k_2,k_3)$ are separable and symmetric products of functions by construction
\begin{equation}\label{eq:Q_primordial}
    Q_{n \leftrightarrow prs}(k_1,k_2,k_3) = \frac{1}{6} \left[ q_p(k_1) q_r(k_2) q_s(k_3) + \mathrm{ perms.} \right] \equiv q_{\{ p} q_{r} q_{s\}}.
\end{equation}
The one-to-one mapping $n \leftrightarrow \{prs\}$ between a basis element index $n$ and its symmetric product of indexed 1D functions is uniquely determined by our choice of basis. The prefactors in \eqref{eq:late-time-inner-product} motivate us to define a separable \textit{CMB shape function}
\begin{equation}
        S_{l_1l_2l_3} \equiv \sqrt{\dfrac{v^2_{l_1}v^2_{l_2}v^2_{l_3}}{C_{l_1}C_{l_2}C_{l_3}}} b_{l_1l_2l_3} = 
        \sum_{n=0}^{n_{\rm max}} \bar{\alpha}_n \bar Q_n (l_1,l_2,l_3),
\end{equation}
where the separable $\bar Q_n(l_1,l_2,l_3)$ is analogously defined via polynomials $\bar q(l_i)$ as in \eqref{eq:Q_primordial}, and the $v_l$ factors are used to separably approximate the geometrical factor $h_{\boldsymbol{L}}$\footnote{$\sqrt{h^2_{l_1l_2l_3}}  \approx v_{l_1}v_{l_2}v_{l_3} \equiv [\left( 2l_1+1 \right)\left( 2l_2+1 \right)\left( 2l_3+1 \right)]^{1/6}$}.
In order to construct separable bases, we must define an inner product $\langle,\rangle_{l/k}$ on the tetrapyd region $\mathcal{V}_{L/k}$:
\begin{equation}\label{inner_product}
\begin{split}
    \langle \Bar{f}, \Bar{g} \rangle_l &= \sum_{l_1,l_2,l_3 \in \mathcal{V}_T} \Bar{w}(l_1,l_2,l_3) \Bar{f}(l_1,l_2,l_3) \Bar{g}(l_1,l_2,l_3)\\
    \langle f, g \rangle_k &= \iiint_{\mathcal{V}_k} w(k_1,k_2,k_3) f(k_1,k_2,k_3) g (k_1,k_2,k_3)  \mathrm{d}\mathcal{V}_k,
    \end{split}
\end{equation}
where the late-time ($l$-space) and primordial ($k$-space) weights, respectively, are given by:
\begin{equation}\label{weight_l}
    \Bar{w}(l_1,l_2,l_3) = \left( \frac{h_{l_1l_2l_3}}{v_{l_1}v_{l_2}v_{l_3}}   \right)^2    
\quad \mathrm{and} \quad
     w(k_1,k_2,k_3) = \frac{1}{k_1+k_2+k_3}.
\end{equation}

The Modal code uses separability in the $k$ and $l$ space to transform the primordial modes of the expansion $\alpha$ into late-time modes $\bar\alpha$ via a projection matrix which contains all the necessary CMB projection effects in Equation \eqref{eq:reduced_bispectrum}. Same approach is applied to the $\tilde B^{(\rm obs.)}$ term in the estimator by separating the triple sum over $l$ and $m$ into a product of \textit{maps}, $\bar{M}_p(\Omega)$, again utilizing the separable mode expansion
\begin{equation}\label{map}
    \bar{M}_p(\Omega) = \sum_{l,m} \bar{q}_p(l) 
    \dfrac{a_{lm}}{v_l\sqrt{C_l}}Y_{lm}(\Omega).
\end{equation}
Integrating these cubic and linear maps over the sphere:
\begin{equation}\label{beta}
    \bar \beta_n = \int \mathrm{d}\Omega \Bar{M}_{p}(\Omega)\Bar{M}_{r}(\Omega)\Bar{M}_{s}(\Omega) 
    -
    \int \mathrm{d}\Omega 
    \langle\Bar{M}^G_{\{p}(\Omega)\Bar{M}^G_{r}(\Omega)
    \rangle
    \Bar{M}_{s\}}(\Omega),
\end{equation}
the estimator simplifies to the ratio of scalar products of vectors in the space of late-time basis functions
\begin{equation}
    \hat f_{\mathrm{NL}} = \frac
    {\bar{\boldsymbol{\alpha}} \cdot \bar{\boldsymbol{\beta}} }
    {\bar{\boldsymbol{\alpha}} \cdot \bar{\boldsymbol{\alpha}} }.
\end{equation}

\subsection{CMB-BEST estimator}
The CMB-BEST (CMB Bispectrum ESTimator) formalism \cite{Sohn:2023fte} is in the Modal family. The primordial bispectrum is first expanded using separable basis functions: the first half of \eqref{eqn:separable_basis_expansion}. Afterwards, however, no basis expansion occurs in the harmonic space. The theoretical bispectrum is computed using the separability of the integral analogously to the KSW estimator:
\begin{align}
	b_{l_1 l_2 l_3} &= \sum_{n \leftrightarrow (p,q,r)} \alpha_{n} \int dr \; \tilde{q}_{p}(l_1, r) \tilde{q}_{q}(l_2, r) \tilde{q}_{r}(l_3, r),  \\
\tilde{q}_{p}(l, r) &\equiv \frac{2r^\frac{2}{3}}{\pi} \int dk \; q_p(k) J_l (kr) T_l (k) .  
\end{align}
From here onwards, the formalism closely resembles that of Modal. The `filtered map' is given by
\begin{align}
	M_p (\Omega, r) \equiv \sum_{l,m} \frac{\tilde{q}_p (l,r)}{C_l} a_{l m} Y_{l m} (\Omega), 
\end{align}
and the cubic and linear terms are computed analogously to Modal except with an additional integral along the line of sight:
\begin{equation}\label{beta}
    \beta_n^{\mathrm{cub}} = \int dr \int \mathrm{d}\Omega M_{(p}(\Omega, r)M_{q}(\Omega,r)M_{r)}(\Omega,r),
    \quad
    \beta_n^{\mathrm{lin}} = \int dr \int \mathrm{d}\Omega 
    \langle M^G_{(p}(\Omega, r)M^G_{q}(\Omega, r)
    \rangle
    M_{r)}(\Omega, r).
\end{equation}

More details on this formalism and its implementation can be found in \cite{Sohn:2023fte,Sohn:2024se}.

\subsection{Look Elsewhere Effect}\label{sec:look-elsewhere}
Constraining primordial non-Gaussianities from models with multiple parameters usually requires extensive search across potentially vast parameter space. Limits of the parameter region are usually put as some priors, containing physically motivated bounds which still give reasonable magnitudes (such as the ratio of sound speeds $c_s$ and mass $\mu$ in this search). We are motivated to search through
one or more parameters $\boldsymbol{\theta}$ to narrow down a parameter subspace which yields the highest signal-to-noise (SNR) ratio 
\begin{equation}
    \sigma_{\rm max} = \max_{\boldsymbol{\theta}} \left| \dfrac{\hat f_{\rm NL}(\boldsymbol{\theta})}{\hat \sigma_{f_{\rm NL}} (\boldsymbol{\theta})} \right|. 
\end{equation}
Higher SNR means this model is better fit to the data, but this result can sometimes be misleading. When interpreting relatively high signal-to-noise ratios for certain parameter choices, one must be careful to distinguish between claiming a genuinely significant detection and a probabilistic outcome from many trials. This \textit{accidental} outcome becomes more likely with the increasing number of observations, a phenomenon commonly known as the \textit{look elsewhere effect}. It is therefore important to have a robust procedure to account for false detections. 

None of the current CMB observations reject the null hypothesis of zero PNGs, when tested against the alternative hypothesis of non-vanishing primordial bispectrum. Standard hypothesis testing goes as follows. The probability, $p$, that $\sigma_{\rm SNR}$ obtains a value greater than some $\sigma_{\rm max}$ is calculated under the null hypothesis and it corresponds to the p-value in statistical hypothesis testing:
\begin{equation}
    p = 1 - \rm Erf \left(\dfrac{\sigma_{\rm max} }{\sqrt{2}}\right).
\end{equation}
A p-value less than a predetermined significance level indicates a statistically significant result, meaning the observed data provide strong evidence against the null hypothesis. For example, to reject the null hypothesis in the particle physics community (i.e. claim new particle discovery), a $5\sigma$ significance is usually required. This roughly corresponds to p-value of 1 in 2 million.

The calculation procedure follows an earlier theoretical approach \cite{Fergusson:2015le} that was applied to official \textit{Planck} PNG results. It computes p-values analytically from the chi-squared distribution function based on the assumptions that maximum signal at each fixed parameter follows a chi distribution with certain degrees of freedom and that SNR measurements of sufficiently uncorrelated templates are independent. This approach works well for standard Feature or SinLog bispectra templates, but has proven to be unsuitable for more complicated shapes such as the cosmological collider.

Hence, we employ the MC computation based on the following approach. Assuming vanishing bispectrum under $H_0(\boldsymbol{\theta})$, the signal-to-noise ratio for each parameter $\boldsymbol{\theta}$ is a random variable distributed as a multivariate normal centered around zero with covariance matrix equal to the late-time shape cosine between two templates
\begin{equation}
    C_{ij} = \dfrac{\langle S(\boldsymbol{\theta_i}), S(\boldsymbol{\theta_j}) \rangle_l}{\sqrt{\langle S(\boldsymbol{\theta_i}), S(\boldsymbol{\theta_i}) \rangle_l}\sqrt{\langle S(\boldsymbol{\theta_j}), S(\boldsymbol{\theta_j}) \rangle_l}}.
\end{equation}
This is analogous to the late-time inner product \eqref{eqn:late_time_corr}. For example, if we search for a cosmological collider signal using a scan of $N$ values for mass $\mu$ and $M$ values for the sound speed ratio $c_s$, the distribution gives a vector $\boldsymbol{\sigma}_{\rm SNR}$ with $N \cdot M$ elements. We then generate a large number (usually $10^5$) of realisations
\begin{equation}
    \boldsymbol{\sigma}_{\rm SNR} \sim \mathcal{N}(\boldsymbol{0}, C)
\end{equation}
and for each realisation we find the maximum SNR as $\sigma_{\rm max}\equiv\mathrm{max}\ \boldsymbol{\sigma}_\mathrm{SNR}$. All such $\sigma_{\rm max}$ are collected and the fraction of those that are greater than or equal to the observed $\sigma_{\max}^{(\rm obs)}$ give a stable estimate of $p(\sigma_{\rm max})$ according to the law of large numbers. Finally, we quote the look-elsewhere-adjusted significance $\tilde\sigma_{\rm SNR}$ which corresponds to the usual language of particle physics detection
\begin{equation}
    \tilde\sigma_{\rm SNR} = \sqrt{2} \rm Erf^{-1} \left(1-p(\sigma_{\rm max}) \right).
\end{equation}

\subsection{Two-pipeline validation on shapes with small sound speeds}
CMB constraints on cosmological collider signal have only been produced recently \cite{Sohn:2024se}. This was carried out using the CMB-BEST pipeline, which had previously been thoroughly tested internally and externally against official \textit{Planck} PNG constraints and Modal results \cite{Sohn:2023fte}. However, Modal, as one of the official pipelines used in \textit{Planck}, has not been applied to this new class of shapes before. Since these pipelines are the only two independent, fully general CMB bispectrum estimators that can handle highly oscillatory shapes, it is important to show their mutually consistent results. Verifying the constraints of two estimators against one another would give stronger validation when exploring new yet unconstrained classes of cosmological collider signals, many of which contain rapid oscillations in the squeezed limit that sometimes prove difficult to resolve numerically.

We revisit the three templates previously described in this category: equilateral collider, low-speed collider, and multi-speed non-Gaussianity. These arise in boost-breaking scenarios where the inflaton $\phi$ and mediator $\sigma$ have distinct sound speeds $c_s$ and $c_\sigma$, leading to novel bispectrum shapes due to multiple sound horizon crossings and clock modulation. For more details, we refer the reader to the original works on these shapes cited below and our previous study \cite{Sohn:2024se}.

We focus on these three templates for several reasons. Firstly, they either depend on a single parameter ($\alpha$ in low speed collider) or on a simple two-parameter family ($\mu$, $\phi^{\rm EC}$ in equilateral collider and $c_1$, $c_2$ in multi-speed shape) that can naturally be condensed into one parameter after normalization. This makes it easier to visualize the constraints and provides clearer intuition when giving a physical explanation of limiting behaviour. Secondly, these shapes have a good physical interpretation of their limiting behaviour in the squeezed and equilateral limits, which offers solid ground to internally test consistency with well-known constraints of standard local and equilateral templates. And finally, unlike other cosmological collider shapes (scalar I and II, spin-1 and spin-2) that are extensively studied and altered in this work, the three mentioned shapes with small sound speeds do not have a modified analogue in the new study. Thus, the two-estimator validation bridges the results from two papers and provides a robust consistency check for future work.

Lastly, we list some aspects where the Modal and CMB-BEST estimators differ:
\begin{itemize}
    \item \textit{Mode functions}. The Modal constraints shown are obtained using Fourier modes\footnote{We also add a singlet of custom modes $Q_1 = \left( \dfrac{k_1k_2}{k_3^2} + \rm perms. \right)$ to better capture local-type divergences.} ($q_p(k)=\sin(\omega_p k+\phi_p)$), while the CMB-BEST ones use Legendre polynomial modes\footnote{Modal also uses Legendre basis, but in the late-time space, again with a handful of custom modes.} ($q_p(k)=P_p \left[2(k-k_\mathrm{min})/(k_\mathrm{max}-k_\mathrm{min})\right]$). Although they show excellent agreement when the modal expansion is accurate, there can be some differences when the expansion is less accurate.
    \item \textit{Mode expansions}. The Modal pipeline involves an extra basis expansion at the late-time bispectrum space compared to CMB-BEST. For extremely oscillatory shapes, modal expansion can break down in a different manner.
    \item \textit{Different $k$ range}. For primordial mode expansion, the observable $k$ range is restricted to a smaller region to allow better mode expansion. CMB-BEST has a slightly higher value of $k_\mathrm{min}$ for this purpose, corresponding to effective $\ell_\mathrm{min}\sim 2.8$, resulting in a slightly smaller $k$-region compared to Modal.
    \item \textit{Background parameters}. Modal uses best-fit cosmology to the \textit{Planck} 2015 results for legacy reasons, while CMB-BEST uses one for the \textit{Planck} 2018 results. However, the background parameters have little effect on the constraints, as shown in \cite{Planck:2015zfm}.
\end{itemize}

\noindent \textbf{Low Speed Collider} This signal arises in scenarios where the inflaton propagates much slower than the massive field ($c_s \ll c_\sigma$). As shown in Refs.~\cite{Jazayeri:2022kjy,Jazayeri:2023xcj}, this bump can be systematically shifted by varying the parameter $\alpha = c_s m/H$, offering a clear phenomenological handle to distinguish low-speed collider signatures:
\be \label{eq:lsc}
 S^{{\rm LSC}}(k_1,k_2,k_3) =  S^{\rm eq}(k_1,k_2,k_3) +
 \frac{k_1^2}{3k_2k_3} \[ 1+  \(\alpha\frac{k_1^2}{k_2k_3}\)^2\]^{-1} + {\rm 2~perms.}~,   
\ee
\begin{figure}[ht]
    \centering
    \includegraphics[width=0.7\linewidth]{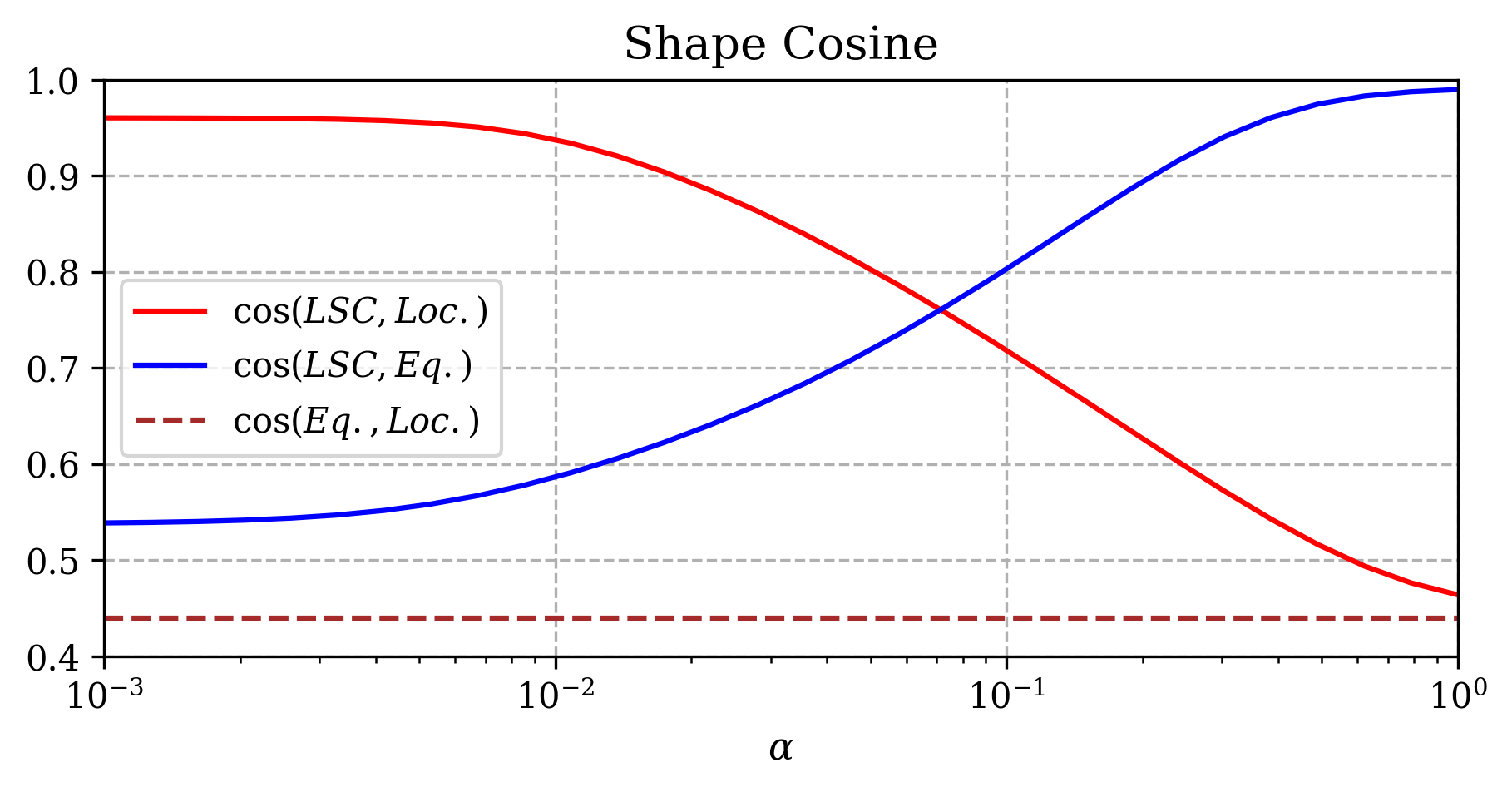}
    \caption{Correlation (shape cosine) between primordial low speed collider and standard shapes: equilateral and local with their mutual correlation added for reference. Note that this plot is slightly different than originally shown in \cite{Jazayeri:2023xcj} because of different definitions of shape cosine, and different weights in the inner product.}
    \label{fig:LSC_corr}
\end{figure}
where $S^{\rm eq}$ is the standard separable template for equilateral shape \eqref{equil} used in Planck analysis. At very low $\alpha$ the signal peaks in the squeezed limit, and as $\alpha$ goes to 1 the signal becomes highly correlated with the equilateral shape, as shown in Figure \ref{fig:LSC_corr}. Joint constraints on its $f_{\rm NL}$ values are shown in Figure \ref{fig:all_three_comparison}. Continuing the analysis in the two limits, at small $\alpha$ we measure central $f_{\rm NL}$ value close to zero with $1\sigma$ confidence levels less than 10, corresponding to the local shape results. At high values of $\alpha$ the error bars grow to the order of $\sim 50$, again consistent with equilateral shape constraints. Furthermore, we observe agreement between the two estimators across the entire parameter region within their $1\sigma$ confidence levels.\\

\noindent \textbf{Multi Speed non-Gaussianity} This bispectrum shape arises from massless exchanges with higher-derivative mixing, where each leg crosses the sound horizon at a different time due to distinct sound speeds. Like the low-speed collider, this leads to resonance enhancements in generic configurations, depending on sound speed ratios. Here we use the following template \cite{Wang:2022eop}:
\be \label{multics}
{S}^{{\rm multi}-c_s} (k_1,k_2,k_3)= \frac{k_1k_2k_3}{(c_1 k_1+ c_2 k_2 +c_3 k_3)^3} +  5~{\rm perms}~, 
\ee
Upon normalizing this template in the equilateral limit, we can without loss of generality fix $c_3=1$ and scan across the other two sound speeds in range $c_1,c_2 \leq c_3$. As one sound speed becomes much greater than the other two (i.e. $c_1,c_2 \rightarrow 0$ in this setup), the shape peaks in the squeezed limit and becomes highly correlated with the local template. As the sound speeds become equal ($c_1,c_2 \simeq c_3 = 1$), the shape approaches equilateral configuration. Figure \ref{fig:all_three_comparison} shows consistent agreement between the two estimates of $f_{\rm NL}$ within their $1\sigma$ confidence levels.\\
\textbf{Equilateral Collider} In standard scenarios, oscillatory features dominate only in the squeezed limit, but with two sound speeds, the squeezed parameter becomes $ u \equiv \frac{c_\sigma k_3}{c_s k_1 + c_s k_2}$, allowing $ u \ll 1 $ even for equilateral configurations when $ c_\sigma \ll c_s $. This leads to oscillatory signals beyond the squeezed limit, giving rise to the so-called equilateral collider shape \cite{Pimentel:2022fsc}. One such template, arising from the interaction $ \dot\phi^2 \sigma $, is
\be \label{eqcol}
 S^{\rm eq.col.}(k_1,k_2,k_3) = \frac{k_1k_2}{(k_1+k_2)^2} \(\frac{k_3}{k_1+k_2}\)^{1/2}\cos\[\mu\log\(\frac{c_\s k_3}{2c_s(k_1+k_2)}\)+\delta^{\rm eq.col.} \] + {\rm 2~perms.}~,
\ee
where $\delta^{\rm eq.col.}$ is a free phase parameter. We note there is a redundancy in parameters $\mu$, $c_{\sigma}$, $c_s$ and $\delta^{\rm eq.col.}$ which can be absorbed into one free phase parameter $\phi^{\rm EC}$, so that the CMB search is performed on the template
\be \label{eqcol_phi}
 S^{\rm eq.col.}(k_1,k_2,k_3) = \frac{k_1k_2}{(k_1+k_2)^2} \(\frac{k_3}{k_1+k_2}\)^{1/2}\cos\[\mu\log\(\frac{k_3}{k_1+k_2}\)+\phi^{\rm EC} \] + {\rm 2~perms.}~
\ee
with suitable normalization in the equilateral limit.

We observe good agreement between two $f_{\rm NL}$ estimators. As shown in Figure \ref{fig:all_three_comparison}, central values lie within $1\sigma$ everywhere except for a small range of intermediate masses where estimates are consistent within $2\sigma$ confidence levels. Since mass of the field has a role of oscillation frequency in the squeezed limit, higher mass signals become increasingly more difficult to faithfully reconstruct. In the Modal case, shape cosine between the theoretical template and its reconstruction are greater than $0.99$ across the entire region, but the total shape correlator\footnote{The total shape correlator $\mathcal{T}$ is defined as
    $1 - \mathcal{T}\left(S_{\rm recon}, S_{\rm th} \right)
    \coloneqq
    \sqrt{\dfrac{\langle S_{\rm recon} - S_{\rm th}, S_{\rm recon} - S_{\rm th} \rangle}{\langle S_{\rm th},S_{\rm th} \rangle}}$.
    One can think of this number as describing the average relative point-wise error of the template reconstruction, and is generally stricter in penalizing poor reconstruction than the simple shape cosine.} drops below $0.9$ for $\mu > 6.8$. This motivates the search for new cosmological collider signals to be limited to lower masses, while also noticing the lack of high significance at higher masses.
\begin{figure}[ht]
    \centering
    \begin{subfigure}[b]{0.49\linewidth}
        \centering
        \includegraphics[width=\linewidth]{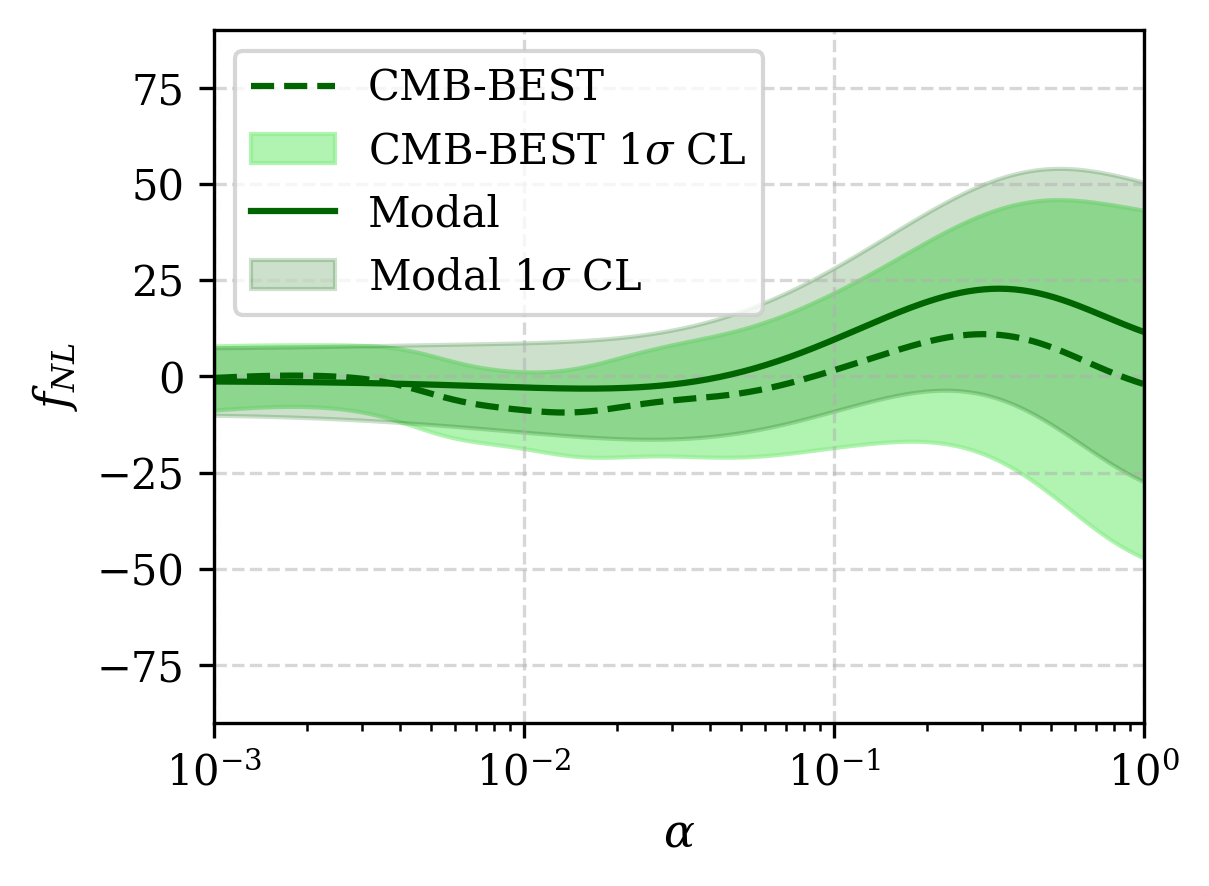}
    \end{subfigure}
    \hfill
    \begin{subfigure}[b]{0.49\linewidth}
        \centering
        \includegraphics[width=\linewidth]{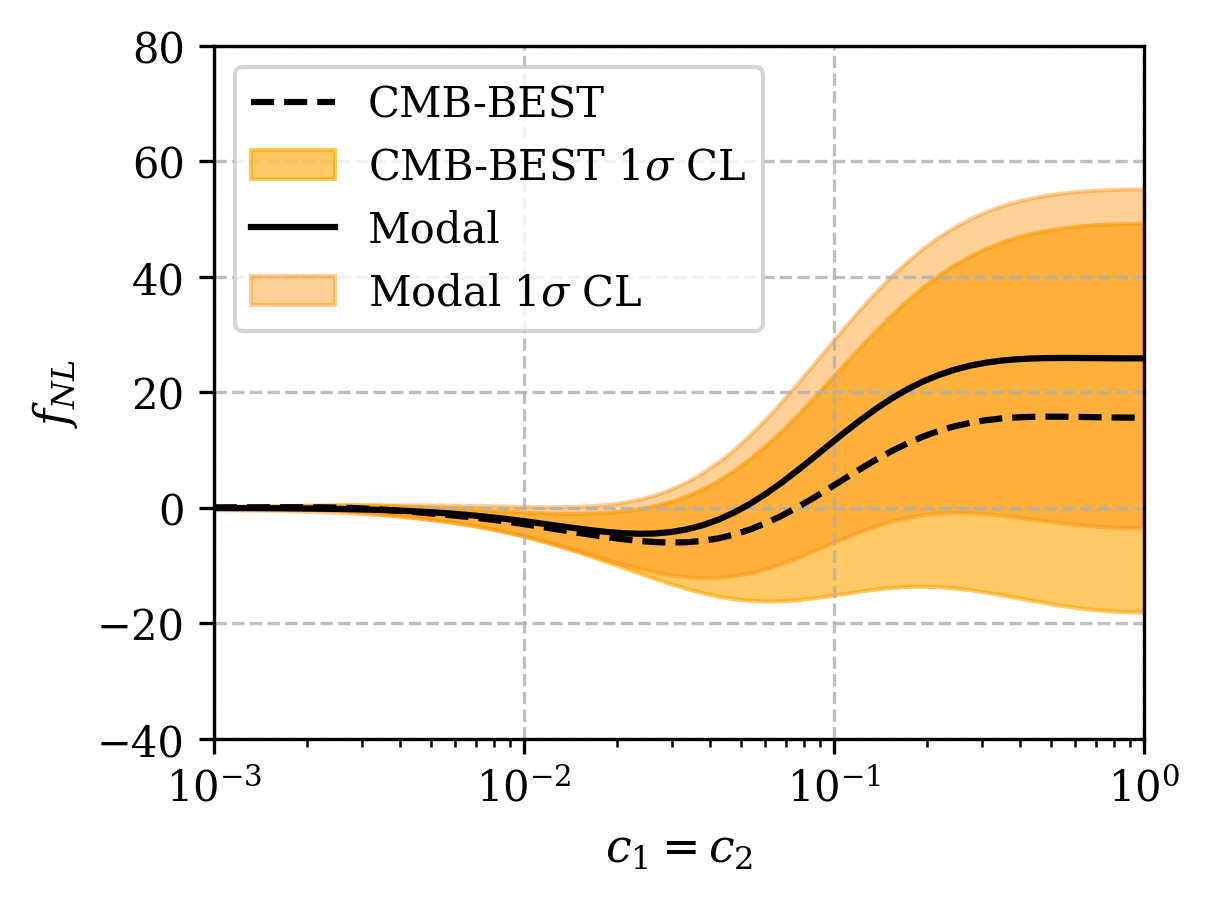}
    \end{subfigure}
    \vspace{1em}
    \begin{subfigure}[b]{0.98\linewidth}
        \centering
        \includegraphics[width=\linewidth]{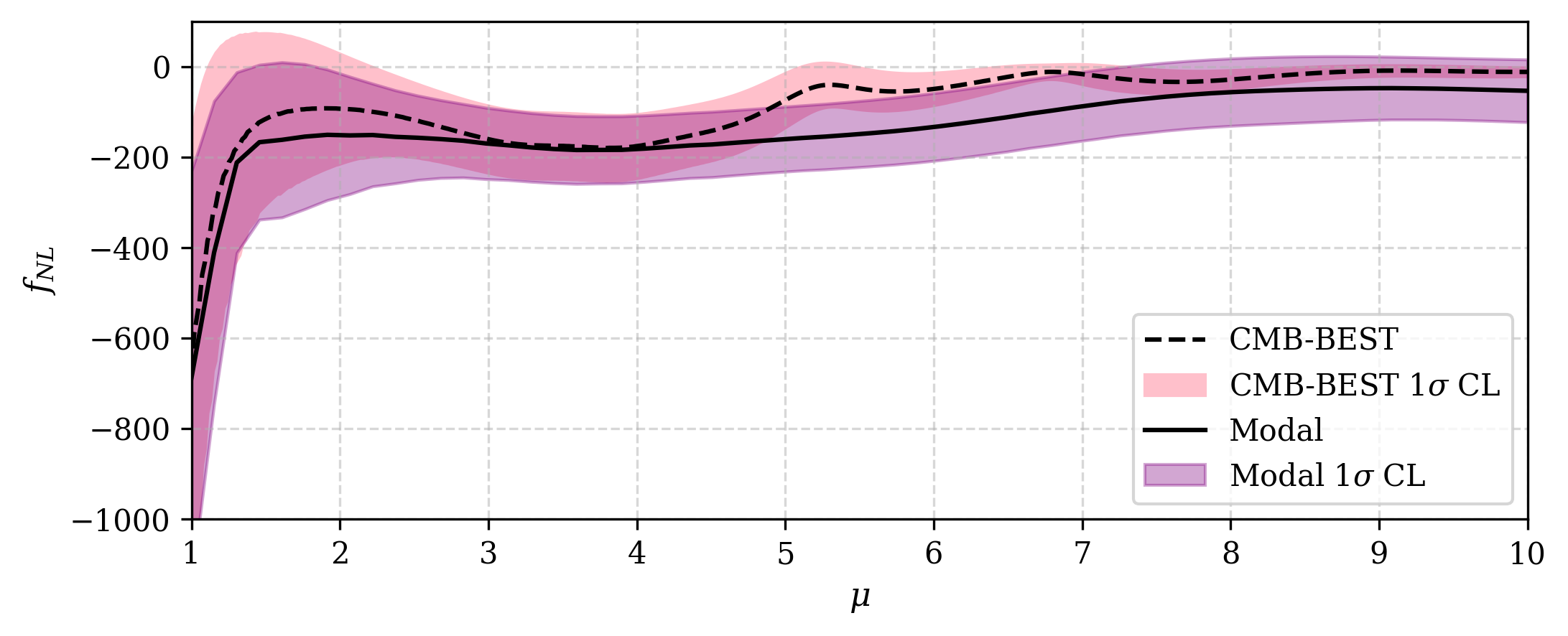}
    \end{subfigure}
    \caption{Two-pipeline validation between Modal and CMB-BEST, applied on three distinct low speed collider shapes from \cite{Sohn:2024se}. We plot central $f_{\rm NL}$ values with their $1\sigma$ error bars ($68\%$ confidence regions). \textit{Top left}: low speed collider \eqref{eq:lsc}. \textit{Top right}: multi speed non-Gaussianity \eqref{multics}. \textit{Bottom}: equilateral collider \eqref{eqcol_phi}. In the first two cases, the two results agree within $1\sigma$ for all parameter values. In the equilateral collider case, the two pipelines produce $f_{\rm NL}$ constraints consistent within $1.5\sigma$ for all masses.}
    \label{fig:all_three_comparison}
\end{figure}

\section{Constraints from Planck 2018 Data}
\label{sec:planck}
In this section, we show $f_{\rm NL}$ constraints for the scalar-I \eqref{scalarI}, scalar-II \eqref{scalarIIa}, spin-1 \eqref{spin1m_simply} and spin-2 \eqref{spin2m} cosmological collider signals using the Modal CMB bispectrum estimator applied to the \textit{Planck} 2018 \texttt{SMICA} $T+E$ data.

In an attempt to normalize the templates in the equilateral limit $S(k_*, k_*, k_*) = 1$, we impose the oscillating $\cos(\mu\log k... )$ term to $1$. Here, $k_* = 0.05 \text{ Mpc} ^{-1}$ denotes a fixed pivot scale, as used in the \textit{Planck} analysis. All collider templates are scale invariant, so this arbitrary choice fixes the normalization uniquely. If we had naively tried to normalize the shape as usual, we would run into divergences every time the cosine became zero. Hence, our stable normalization removes the exponentially suppressed signal at higher mass that occurs from the $\sinh(\mu)$ and similar terms, and it additionally removes the $\mu$-dependent amplitude from $\beta$. This brings magnitudes of the entire class of collider shapes to an equal footing when it comes to comparing $f_{\rm NL}$ constraints for different values of $\mu$ even though the equilateral limit of some templates would not necessarily equal to unity. With such normalization, direct comparison of the Modal reconstruction quality becomes straightforward, where now the only challenge is posed by the oscillation frequency, and not by the exponentially suppressed values below machine precision error. However, we point out that for some values of $\mu$ and $c_s$ the cosine term in the equilateral limit might deviate significantly from unity. For this reason, and for the later computation of the look elesewhere effect (as described in \ref{sec:look-elsewhere}), we mainly focus on the signal to noise ratio $\hat f_{\rm NL} / \hat \sigma_{f_{\rm NL}}$, which is independent of the normalization choice, as the relevant metric when assessing the presence of collider templates in the CMB data.

We use full $160$ simulated Gaussian ($f^{G, sim}_{\rm NL} = 0$) \textit{Planck} maps to compute the linear term in $\bar\beta_n$ in Equation \eqref{beta}, and to estimate the uncertainties in the $f_{\rm NL}$ constraints. All error bars on $f_{\rm NL}$ constraints that we quote correspond to $68\%$ confidence level.

All templates and their $f_{\rm NL}$ constraints depend on the same set of parameters $(\mu, c_s)$. The mass parameter of the exchange field, $\mu$, is scanned uniformly between 1 and 6. We limit our search up to $\mu=6$ as high mass parameter makes their highly oscillatory bispectrum shapes more challenging to accurately reconstruct. Even so, CMB constraints in the earlier work \cite{Sohn:2024se} and our preliminary run of the pipeline at higher $\mu$s did not find significant non-Gaussianities that would require reevaluation of the ability to faithfully reconstruct and search for high-mass signals. The ratio of sound speeds, in this work described by a single variable $c_s$, is the ratio between the sound speed of an exchange field and the sound speed of the inflaton. This means that it can take values across a vast range of orders of magnitude. Our search is carried out in the log-uniform\footnote{Although shapes depend on natural logarithm of the sound speed, we scan across equally spaced powers of 10 in sound speed, i.e. in $\log_{10}$.} range between $c_s = 10^{-2}$ (sound speed of the exchange field much lower than that of the inflaton) and $10^{2}$ (sound speed of the exchange field much greater than that of the inflaton).

\subsection{Collider signals from massive scalars}
In this subsection, we focus on the oscillations in the squeezed limit from the massive scalar exchange in Section \ref{sec:scalar-osc}. We provide constraints on the (normalized) shapes \eqref{scalarI} and \eqref{scalarIIa}, as well as their orthogonalized counterparts.

We present the results as follows. The significance, defined as the absolute signal-to-noise ratio, is shown as a two-dimensional heat map in the $(\mu, c_s)$ plane. Below this, we display the significance maximized over $c_s$. Specifically, for each fixed value of $\mu$, we determine the value of $c_s$ that maximizes the absolute signal-to-noise ratio, yielding a smooth curve $\sigma_{\rm SNR}^{\rm max} (\mu)$.
One could do the opposite, maximizing over $\mu$ at fixed $c_s$. This procedure yields a highly oscillatory SNR curve with no clear physical interpretation. In the context of particle detection, it is more natural to search for signal enhancements as a function of the mass parameter, which directly probes characteristic energy scales. By contrast, the effective sound speed ratio is subject to intrinsic model degeneracies and is therefore less physically informative.\\
\begin{figure}[ht]
    \centering
    \begin{subfigure}[b]{0.49\linewidth}
        \centering
        \includegraphics[width=\linewidth]{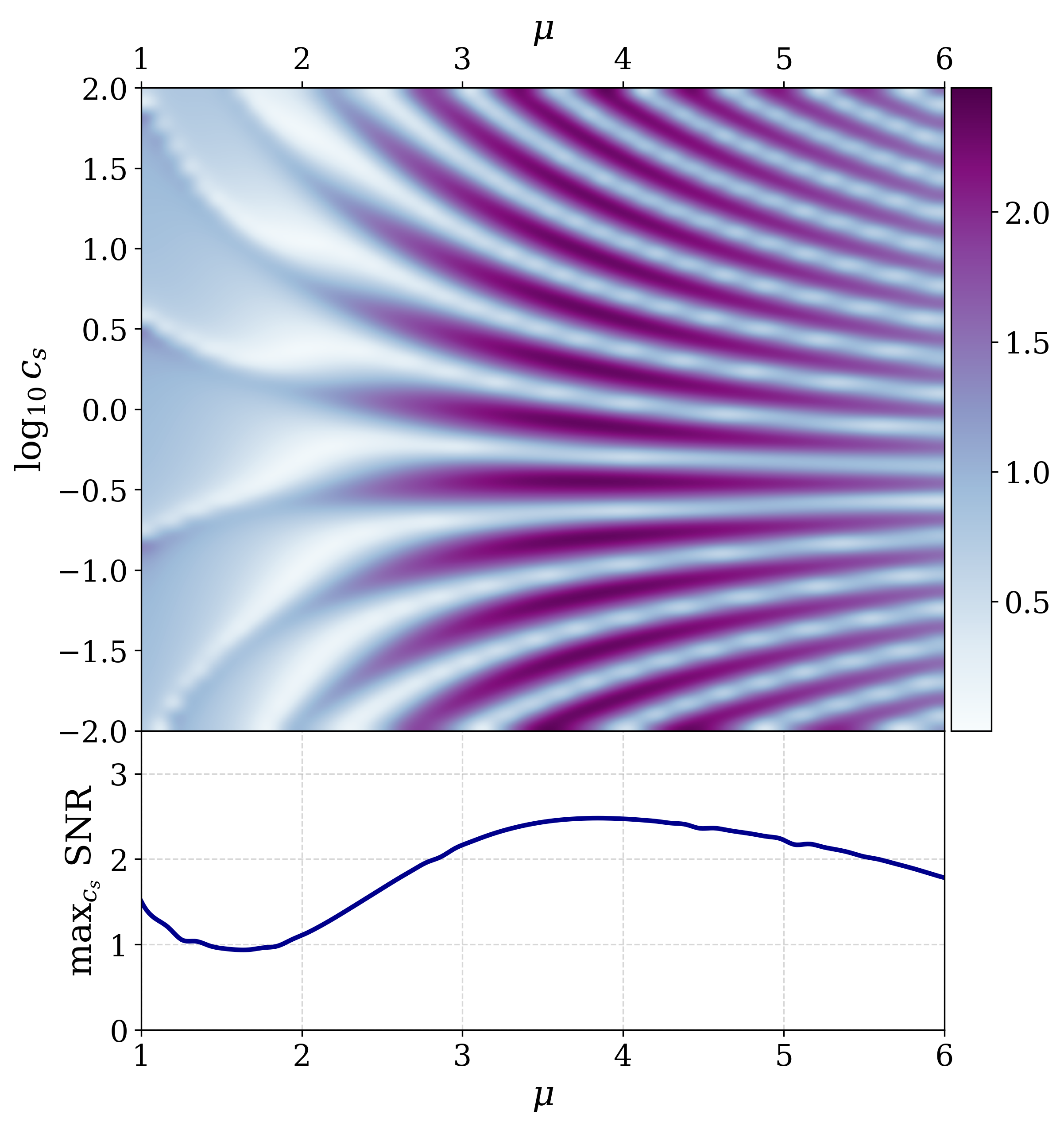}
    \end{subfigure}
    \hfill
    \begin{subfigure}[b]{0.49\linewidth}
        \centering
        \includegraphics[width=\linewidth]{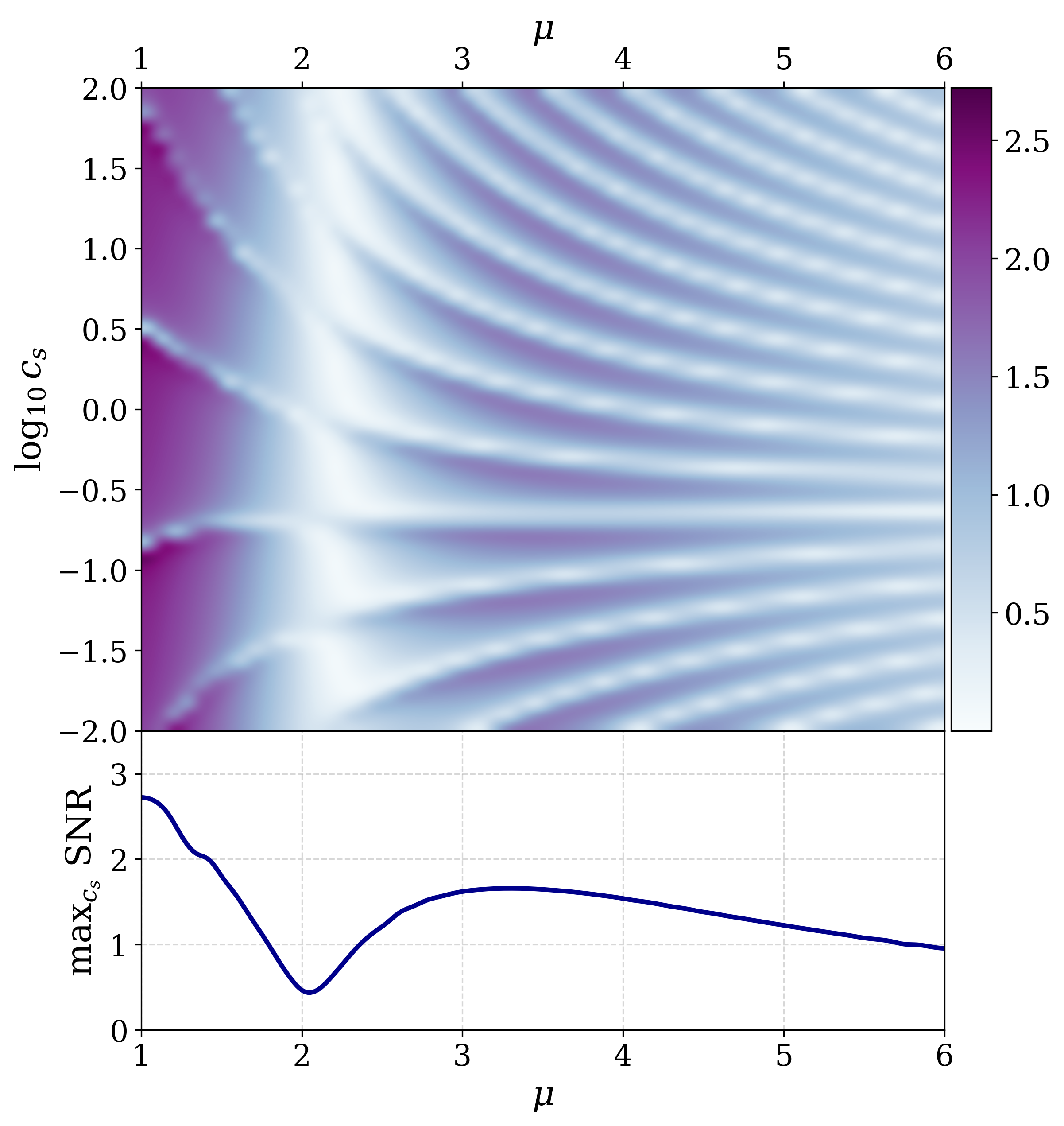}
    \end{subfigure}
    \caption{Signal to noise ratio for the scalar-I collider \eqref{scalarI} (left) and its orthogonalized counterpart (right), as a function of the mass parameter $\mu$ and the sound speed ratio $c_s$. The maximum significance of $2.48$ is obtained for $(\mu, c_s) = (3.88, 45.82)$ in the former case and the significance of $2.72$ for $(\mu, c_s) = (1.00, 0.12)$ in the latter.}
    \label{fig:scalar-1_snr}
\end{figure}

\begin{figure}[ht]
    \centering
    \begin{subfigure}[b]{0.49\linewidth}
        \centering
        \includegraphics[width=\linewidth]{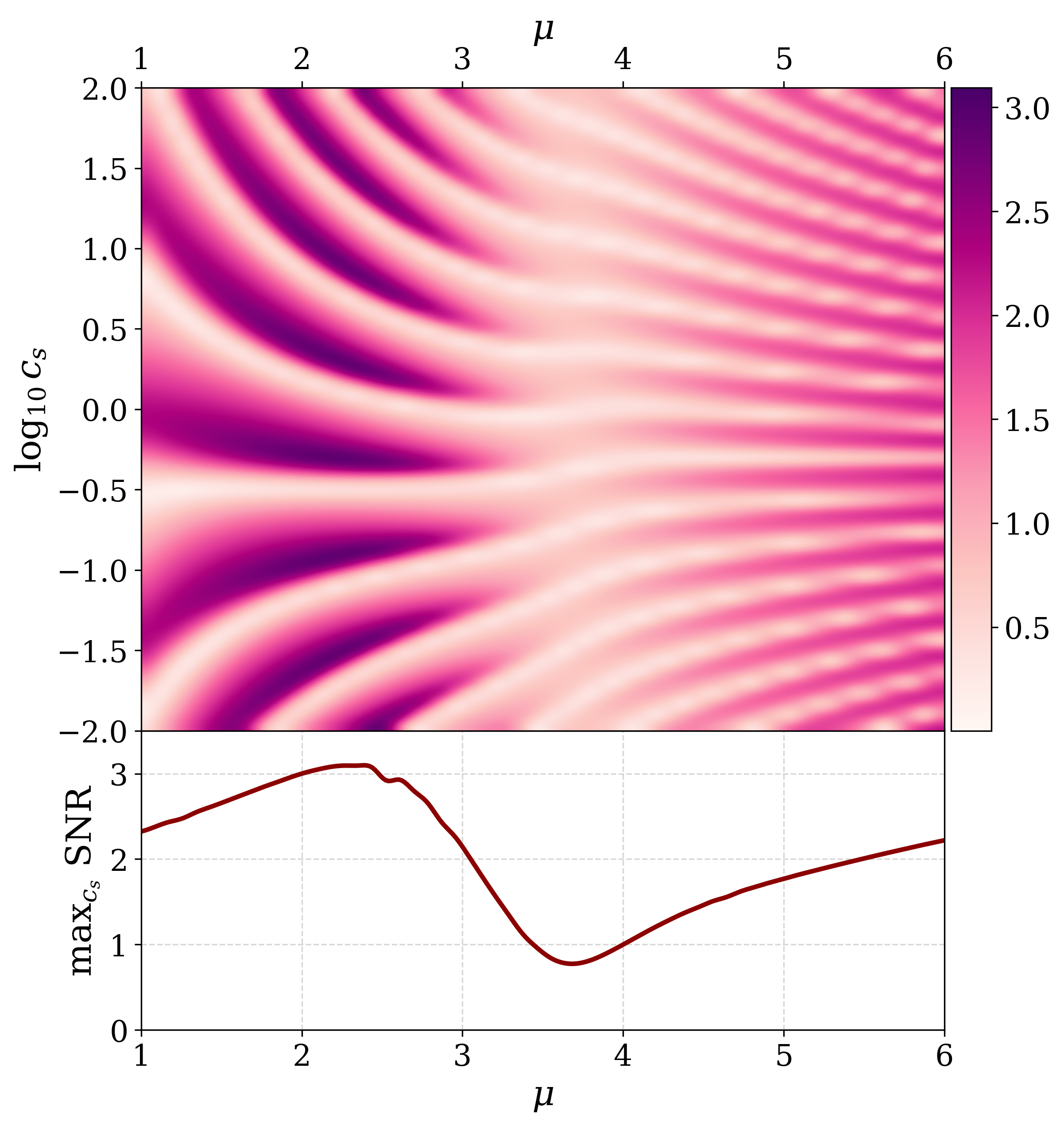}
    \end{subfigure}
    \hfill
    \begin{subfigure}[b]{0.49\linewidth}
        \centering
        \includegraphics[width=\linewidth]{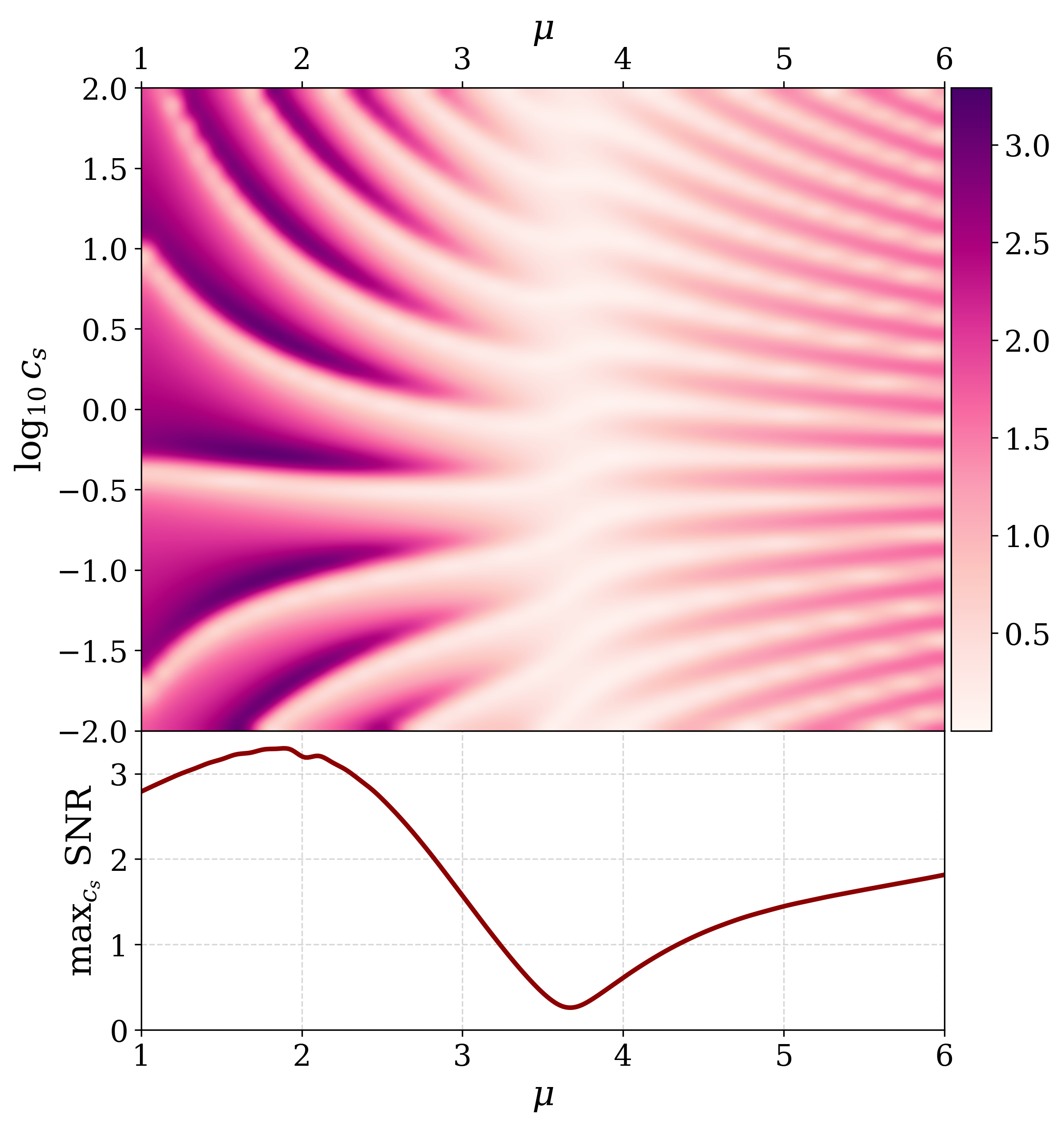}
    \end{subfigure}
    \caption{Signal to noise ratio for the scalar-II collider \eqref{scalarIIa} (left) and its orthogonalized version (right), as a function of the mass parameter $\mu$ and the sound speed ratio $c_s$. We detect maximum significance of $3.09$ for $(\mu, c_s) = (2.36, 0.12)$ in the former case and the significance of $3.29$ for $(\mu, c_s) = (1.85, 0.09)$ in the latter.}
    \label{fig:scalar-2_snr}
\end{figure}

\noindent \textbf{Scalar-I } Figure \ref{fig:scalar-1_snr} shows significance for the scalar-I template \eqref{scalarI}. We note a peak in significance around $\mu \approx 4$. The scalar-I template obtains best-fit constraint of $f_{\rm NL} = 182 \pm 73$ at $\mu = 3.88$ and $c_s = 45.82$. This corresponds to the maximum raw significance of $\sigma_{\rm SNR}^{\rm max} = 2.48$. The orthogonalized version of this template enhances the significance for very light mass fields, but reduces the significance for any $\mu > 3$ compared to the unmodified template, suggesting that the original template was enhanced by a high correlation with the orthogonal template. Best fit constraint of $f_{\rm NL} = 3660 \pm 1310$ is measured at the lower boundary of $\mu \lesssim 1$ and $c_s = 0.12$, corresponding to the maximum raw significance of $\sigma_{\rm SNR}^{\rm max} = 2.72$. Upon adjusting for the look elsewhere effect, the scalar-I template has the maximum significance of $\tilde \sigma_{\rm SNR} = 1.53$ (unorthogonalized) and $\tilde \sigma_{\rm SNR} = 1.71$ (orthogonalized).\\

\noindent \textbf{Scalar-II } Figure \ref{fig:scalar-2_snr} shows significance for the scalar-II template \eqref{scalarIIa}. We report a strong ($\sigma_{\rm SNR}\sim 3$) signal around $\mu \approx 2$. Best fit constraint of $f_{\rm NL} = -295 \pm 95$ is obtained for $\mu = 2.36$ and $c_s = 0.12$. This corresponds to the maximum raw significance of $\sigma_{\rm SNR}^{\rm max} = 3.09$.
Orthogonalization of the scalar-II shape reduces the significance at high mass, while preserving the overall features and even enhancing the signal for $\mu \sim 2$. Consequently, best-fit constraint of $f_{\rm NL} = -600 \pm 182$ is estimated for $\mu = 1.85$ and $c_s = 0.09$, corresponding to the maximum raw significance of $\sigma_{\rm SNR}^{\rm max} = 3.29$.
Adjusted for the look elsewhere effect, the scalar-II shape has the maximum adjusted significance $\tilde \sigma_{\rm SNR} = 2.14$, which then grows to $\tilde \sigma_{\rm SNR} = 2.35$ when orthogonalized.
\subsection{Collider signals from spinning fields}
In this subsection we focus on the angular dependence of the signal from the spinning exchanges in Section \ref{sec:spin}. We provide constraints on the (normalized) shapes \eqref{spin1m_simply} and \eqref{spin2m}, as well as their orthogonalized counterparts.\\
\begin{figure}[ht]
    \centering
    \begin{subfigure}[b]{0.49\linewidth}
        \centering
        \includegraphics[width=\linewidth]{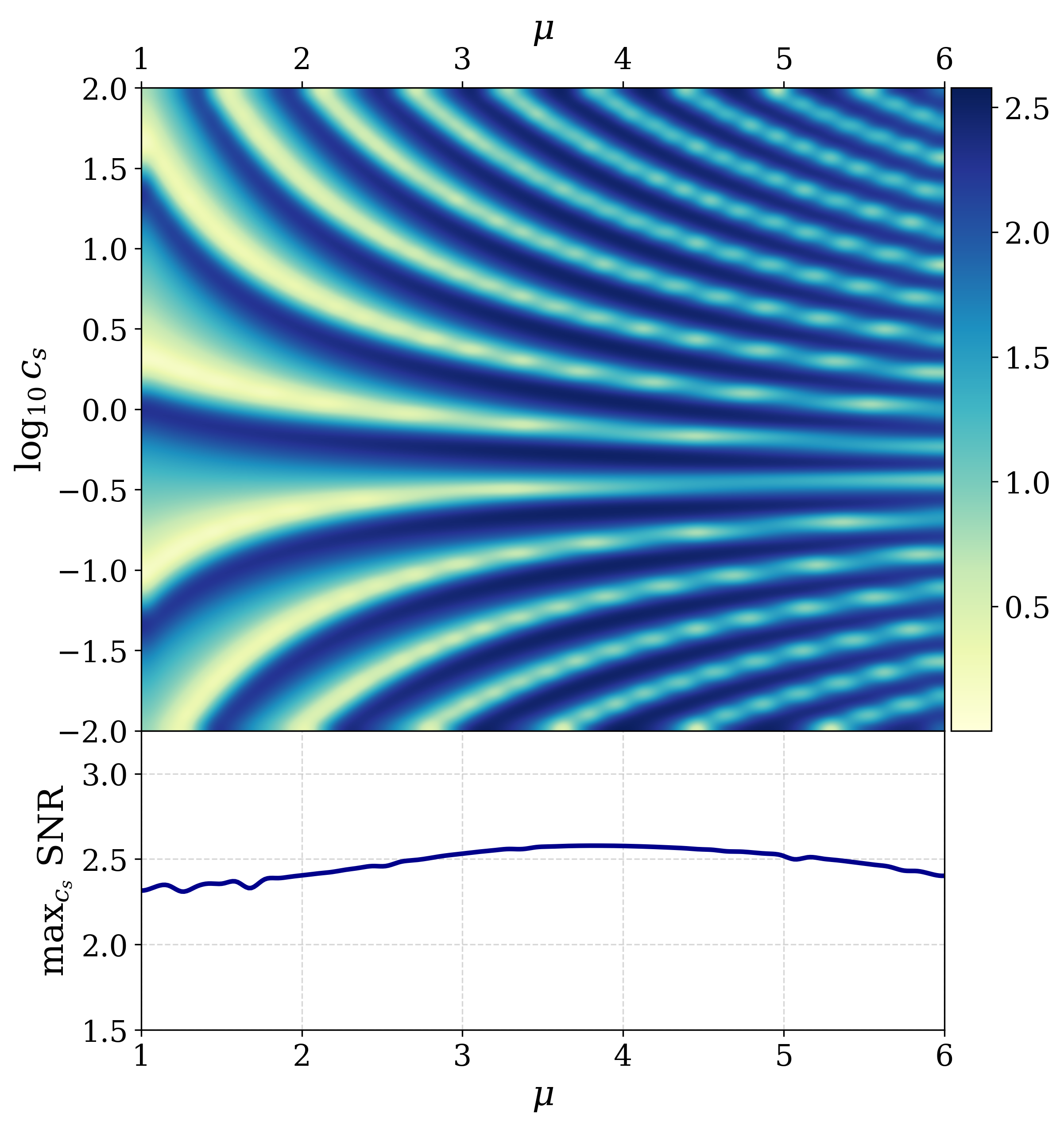}
    \end{subfigure}
    \hfill
    \begin{subfigure}[b]{0.49\linewidth}
        \centering
        \includegraphics[width=\linewidth]{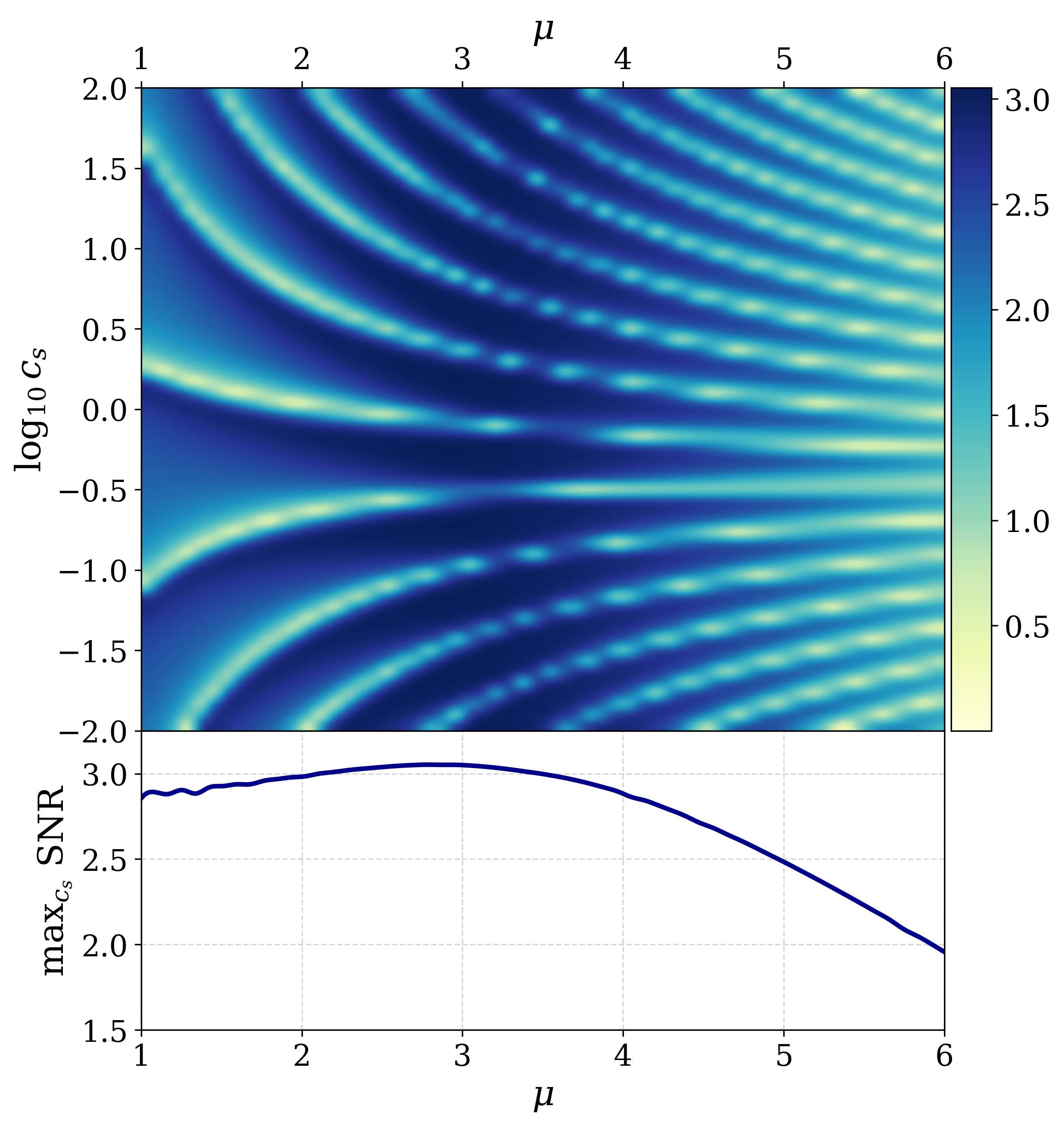}
    \end{subfigure}
    \caption{Signal to noise ratio for the spin-1 collider \eqref{spin1m_simply} (left) and its orthogonalized counterpart (right), as a function of the mass parameter $\mu$ and the sound speed ratio $c_s$. The maximum significance of $2.58$ is obtained for $(\mu, c_s) = (3.80, 2.76)$ in the former case, and in the latter case we detect the significance of $3.05$ for $(\mu, c_s) = (2.78, 1.73)$ in the latter.}
    \label{fig:spin-1_snr}
\end{figure}

\noindent \textbf{Spin-1} Figure \ref{fig:spin-1_snr} shows the significance for the spin-1 template \eqref{spin1m_simply}. The signal shows a very flat SNR curve, peaking for $\mu = 3.80$ and $c_s = 2.76$ with $f_{\rm NL} = -132 \pm 51$. This corresponds to the maximum raw significance of $\sigma_{\rm SNR}^{\rm max} = 2.58$. This weak $f_{\rm NL}$ dependence on the exchange mass can be explained by the fact that the odd-spin collider signals are more suppressed relative to the EFT contribution, rendering the SNR largely insensitive to the mass parameter \cite{Pimentel:2022fsc, Lee:2016vti}. 
Orthogonalizing the spin-1 template removes this EFT background, revealing more features in the SNR curve. Notably, the highest signal falls in the high mass region, while the significance is now the highest for intermediate masses around $\mu \sim 3$. The best-fit constraint of $f_{\rm NL} = 483 \pm 158$ is measured for $\mu = 2.78$ and $c_s = 1.73$, corresponding to the maximum raw significance of $\sigma_{\rm SNR}^{\rm max} = 3.05$. The unorthogonalized and orthogonalized spin-1 shapes have an adjusted significance of $\tilde \sigma_{\rm SNR} = 1.86$ and $\tilde \sigma_{\rm SNR} = 2.14$, respectively, after accounting for the look elsewhere effect.\\
\begin{figure}[ht]
    \centering
    \begin{subfigure}[b]{0.49\linewidth}
        \centering
        \includegraphics[width=\linewidth]{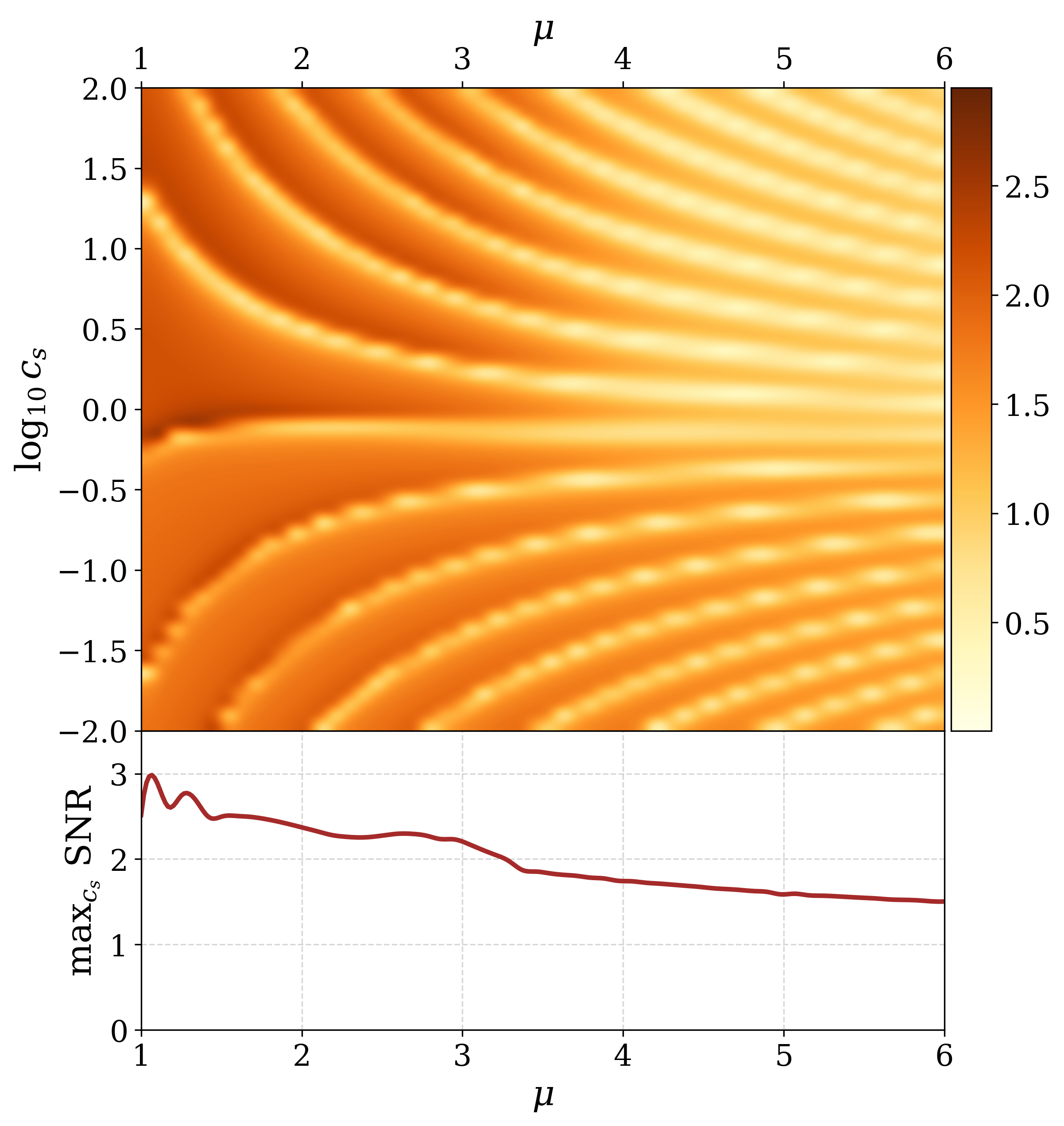}
    \end{subfigure}
    \hfill
    \begin{subfigure}[b]{0.49\linewidth}
        \centering
        \includegraphics[width=\linewidth]{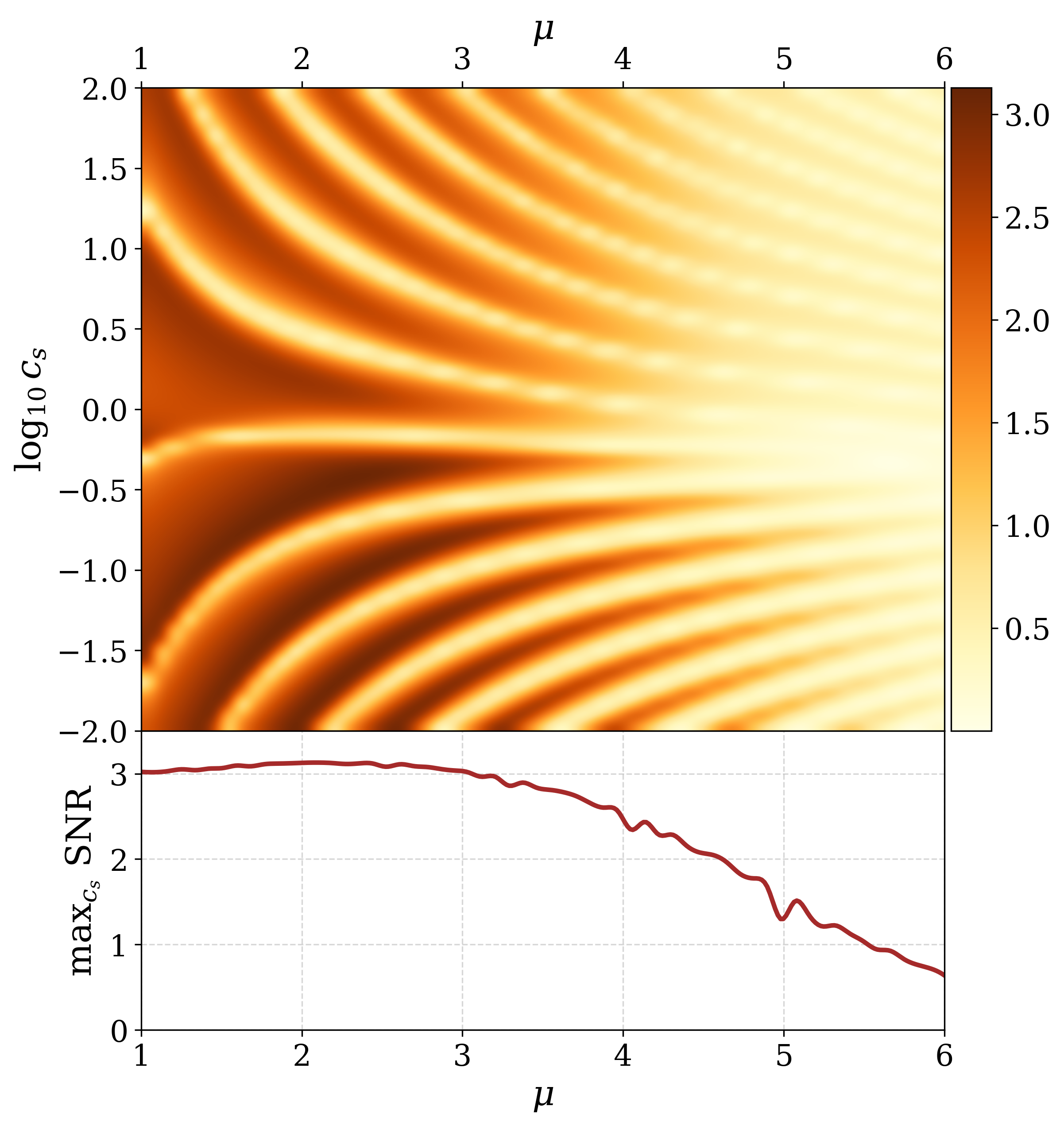}
    \end{subfigure}
    \caption{Signal to noise ratio for the spin-2 collider \eqref{spin2m} (left) and its orthogonalized shape (right), as a function of the mass parameter $\mu$ and the sound speed ratio $c_s$. The maximum significance of $2.95$ is obtained for $(\mu, c_s) = (1.09, 0.68)$ in the former case and the significance of $3.13$ for $(\mu, c_s) = (2.10, 0.07)$ in the latter.}
    \label{fig:spin-2_snr}
\end{figure}

\noindent \textbf{Spin-2} Figure \ref{fig:spin-2_snr} shows the significance for the spin-2 template \eqref{spin2m}. We observe the strongest signal at lower masses $(\mu \approx 1)$ with a steady drop beyond $\mu > 3$. The best-fit constraint of $f_{\rm NL} = 19 \pm 6.5$ is observed for $\mu = 1.09$ and $c_s = 0.68$. This corresponds to the maximum raw significance of $\sigma_{\rm SNR}^{\rm max} = 2.95$. 
Orthogonalizing this template increases the significance of the signal, with the best-fit constraint of $f_{\rm NL} = 70 \pm 22$ being observed for $\mu = 2.10$ and $c_s = 0.07$, corresponding to the maximum raw significance of $\sigma_{\rm SNR}^{\rm max} = 3.13$. Accounting for the look elsewhere effect, the spin-2 unorthogonalized and orthogonalized shapes obtain maximum adjusted significance of $\tilde \sigma_{\rm SNR} = 1.93$ and $\tilde \sigma_{\rm SNR} = 1.97$, respectively.

The spin-2 template is the only one considered in this paper that evidently shows asymmetry in $f_{\rm NL}$ constraints with respect to the $\log c_s = 0$ line. This phenomenon is enhanced at higher $\mu$'s. It can be explained by investigating \eqref{spin2m} which contains $c_s$ not only within the $\cos\log$ structure, but also as different power factors in the product. If we treat the template as the sum of two terms (up to permutations), we note that the first term scales as $c_s^{-1/2}$, while the second term scales as $c_s^{-5/2}$. This means that the normalization scheme does not cancel the $c_s$ dependence of the amplitude, resulting in this asymmetry. One can easily check that this feature is not present in other templates.

\begin{figure}
    \centering
    \includegraphics[width=0.5\linewidth]{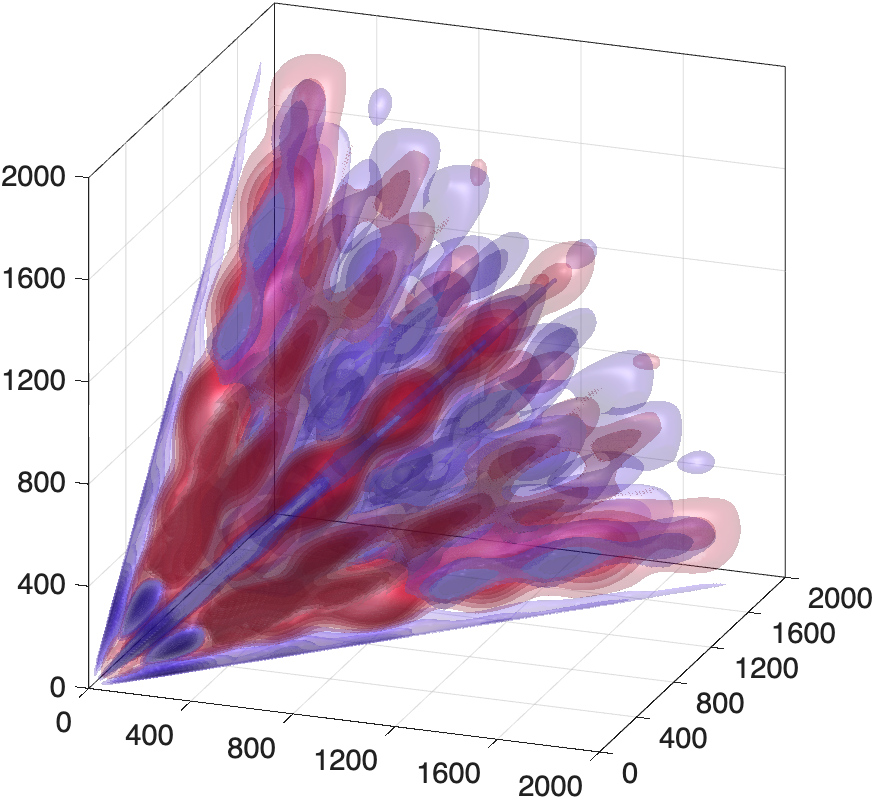}
    \caption{CMB bispectrum (normalized with respect to the Sachs-Wolfe constant model) of the most significant orthogonalized scalar-II template with $\mu = 1.85$ and $c_s = 0.09$. The corresponding primordial bispectrum is presented in Fig. 2 of the companion paper \cite{Suman:2025vuf}. Although a variety of bispectrum estimators exist for analyzing primordial shapes, the Modal approach is unique in its ability to visualize and systematically manipulate the CMB bispectrum through its late-time separable basis, a feature not available in other pipelines.}
    \label{fig:best-fit_latetime_bispectrum_recon}
\end{figure}

\section{Summary and Conclusions}
\label{sec:concl}

In this work, we have presented an updated and systematic search for cosmological collider signatures of primordial non-Gaussianity (PNG) in the latest \textit{Planck} CMB data \cite{Akrami:2019izv}, employing the Modal bispectrum estimator \cite{Fergusson:2009nv, Fergusson:2010dm, Liguori:2010primordial, Fergusson:2014gea}. Building on earlier efforts \cite{Sohn:2024se}, we provide a comprehensive study of a set of cosmological collider templates. Furthermore, we develop a framework for isolating signals of massive particle exchange from the single-field EFT background. The orthogonalized templates are developed for an extensive parameter search using \textit{Planck} CMB, and crucially, they can be applied to all future CMB and LSS surveys with minimal modification. The main results and their implications are summarized below.

First, under the minimal theoretical assumptions of approximate scale invariance and weak couplings, we developed a full basis of cosmological collider templates, spanning a wide range of particle masses and spins. We note that, in general, one can separate the primordial bispectrum shape of a single-exchange process into a sum of an EFT background and the collider signature of massive particle production. While the former resembles single field inflation (producing equilateral- and orthogonal-like templates), the latter corresponds to a unique massive particle signal which cannot be mimicked by any other standard templates. Such collider shapes show intrinsic oscillations in the squeezed limit and depend on two parameters: mass of the exchange field $\mu$ and the sound speed ratio between the massive field and the inflaton $c_s$. By considering quadratic and cubic interactions, we lay out a set of collider shapes: two scalar exchange templates, and one for each spin-1 and spin-2 fields.

It was shown that such collider templates can still exhibit high correlation with the standard templates, but its single-field component can be absorbed into the EFT background, allowing us to construct a full linear combination of these two components. This key aspect of our approach allows systematic orthogonalization of these templates with respect to single-field inflationary PNG shapes, thereby removing degeneracies that would otherwise obscure the interpretation of observational constraints. As discussed in detail, both marginalization and orthogonalization offer viable pathways to isolating signatures of massive particles; however, we have adopted the orthogonalization strategy in order to obtain constraints that are both robust and easily comparable across different surveys. This methodology represents an essential step toward a statistically meaningful detection of cosmological collider signals and enables a clear separation between genuinely new physics and contributions consistent with single-field inflation.

Secondly, we applied the Modal estimator to search for given PNGs in the CMB data, using the \textit{Planck} 2018 legacy release with both temperature and polarization anisotropies. Since the previous work used CMB-BEST pipeline, we internally tested against the two estimators and obtained consistent results within uncertainties. This adds another layer of consistency checks on top of previously compared performance for standard templates \cite{Sohn:2023fte}. Applying this analysis to a vast parameter space of collider templates, we have identified the most significant indication of massive particle exchange in the CMB data so far. The strongest signal corresponds to an orthogonalized massive scalar exchange template \eqref{scalarIIa} with mass parameter $\mu = 1.85$, yielding a significance of $2.35\sigma$, adjusted for the look-elsewhere effect. This result is consistent with the previous constraints obtained with CMB-BEST, where the Scalar-II template also showed the strongest signal ($1.8\sigma$ adjusted), and for similar mass ($\mu = 2.13$). Although the template has been modified and improved since then, it still captures the same physical processes. This best-fit reconstructed (late-time) CMB bispectrum is plotted in figure \ref{fig:best-fit_latetime_bispectrum_recon} (for temperature only). It is illuminating to compare this with the projected CMB equilateral, orthogonal, DBI and other shapes illustrated in ref.~\cite{Fergusson:2006pr} and then the actual full Modal bispectrum reconstruction shown in Figure 10 of the Planck 2018 NG paper \cite{Akrami:2019izv}.  These show the equilateral bispectrum modified by orthogonal and local signals contributing to the predicted bispectrum as expected, and some match to the measured Planck CMB bispectrum signal.

Other noteworthy signals have been obtained from orthogonalized spin-1 and spin-2 fields, with $2.14\sigma$ and $1.98\sigma$ results, respectively, after adjusting for the look elsewhere effect. Orthogonalization of the collider templates removes the well-constrained single-field EFT background which explains the enhancement of the signal.

It was recently shown \cite{Coulton:2022wln} that PNGs from standard single field templates might be biased by secondary CMB anisotropies, so finding a significant collider signal, uncorrelated with these templates, would add confidence in detecting genuine primordial non-Gaussianities from massive exchange. 
While this result does not yet constitute a detection, it provides an intriguing hint of a non-zero scalar bispectrum beyond the single-field paradigm. Should such a signal be enhanced by future, more precise observations, it would have profound implications for our understanding of inflation, effectively ruling out the simplest single-field models and pointing toward a richer particle spectrum active during the inflationary epoch.

It would be instructive to see the application of a new independent CMB bispectrum estimator \cite{Philcox:2025bbo} on this set of collider templates across the given parameter space. Beyond the specific results obtained with \textit{Planck}, an important outcome of this work is the generality and flexibility of the data-analysis strategy. The orthogonalized template approach developed here can be readily applied to forthcoming CMB datasets, including those from the Simons Observatory \cite{ade2019simons}, which are expected to substantially improve sensitivity to oscillatory PNG signals. Moreover, the same methodology can be extended to higher-order correlators, such as the primordial trispectrum, where signals from spinning particles are expected to exhibit even richer angular and momentum-space structures. Recent progress in this direction has been reported in \cite{Philcox:2025bvj,Philcox:2025lrr,Philcox:2025wts}, and our framework provides a natural starting point for systematic searches in this sector. The Modal framework already has the trispectrum estimator implemented as another part of its pipeline \cite{Fergusson:2010gn, Regan:2010cn} and can be applied to such templates in our future work.

Looking further ahead, LSS surveys will play an increasingly important role in testing cosmological collider physics. Upcoming experiments such as SphereX, Euclid, and LSST are expected to deliver high-precision measurements of galaxy clustering over a wide range of scales and redshifts, offering complementary probes of PNGs \cite{MoradinezhadDizgah:2017szk,MoradinezhadDizgah:2018ssw,Goldstein:2025eyj}. In particular, recent forecast studies \cite{Anbajagane:2025uro} indicate that LSS observations may achieve constraints on collider PNG that are competitive with, or even surpass, those obtained from current CMB analyses \cite{Sohn:2024se}. It will be highly informative to assess whether the use of more accurate and more general collider templates, such as those developed in this work, can further strengthen these forecasts.

Finally, the framework developed here also bears on the interpretation of the existing
single-field constraints. Since the collider templates exhibit a non-zero correlation with the equilateral and orthogonal shapes, a massive-exchange contribution left
unaccounted for in a single-field analysis shifts the constraints by
$\Delta f_{\rm NL}^{\rm equi}/\sigma(f_{\rm NL}^{\rm equi}) =
\text{SNR}_{(\rm coll)}\cdot\text{Corr}(\text{equi,coll})$, and likewise for the orthogonal
shape, where the shape correlation term is computed in the CMB $\ell$-space. This bias is linear in the collider amplitude, and
across the full parameter scan we find shifts $\lesssim 1\sigma$ in $f_{\rm NL}^{\rm equi}$ and $\lesssim 2\sigma$ in $f_{\rm NL}^{\rm ortho}$. Hence, the \textit{Planck}
single-field constraints are to be understood as conditional on the absence of collider
signals, while shifts of the order given above could occur in the presence of massive exchange. 
If several massive species contribute, as in the general
expansion~(2.10), the induced shifts may partially cancel among themselves and against the
genuine single-field signal. Conversely,
wherever the implied shift is large, a collider interpretation would require the single-field
amplitudes to take values in tension with the measured $f_{\rm NL}^{\rm equi}$ and
$f_{\rm NL}^{\rm ortho}$. Therefore, consistency with the
\textit{Planck} single-field constraints could provide an independent check on any claimed collider
signal. We leave a detailed analysis with joint
constraints on the single-field and multiple collider shapes for future work.

In summary, we have performed a systematic and model-independent search for cosmological collider signatures in the \textit{Planck} CMB data, combining a complete set of collider templates with an orthogonalization strategy and a robust Modal bispectrum estimator. Our results yield the most significant indications of massive particle exchange reported so far, with consistent outcomes across independent analysis pipelines and removed degeneracies with single-field inflationary backgrounds. While no definitive detection is yet achieved, the emerging preference for scalar exchange provides a clear target for future observations. The framework developed here is readily extendable to upcoming CMB and LSS surveys, offering a unified approach to probing the particle content of the inflationary Universe.

\vskip16pt
\paragraph{Acknowledgements} 

We would like to thank Santiago Agüi Salcedo, Giovanni Cabass, Xingang Chen, Ciaran McCulloch, Xi Tong, Bowei Zhang for inspiring discussions. We also thank the anonymous referee for constructive comments. JF and PS are supported by Science and Technology Facilities Council (STFC) grant ST/X508287/1. DGW is partially supported by a Rubicon Postdoctoral Fellowship awarded by the Netherlands Organisation for Scientific Research (NWO), the Stephen Hawking Centre for Theoretical Cosmology, and GRF project 16306425 from the Research Grants Council of Hong Kong. WS is supported by the SCIPOL project funded by the European Research Council (ERC) under the European Union’s Horizon 2020 research and innovation program (Grant agreement No. 101044073).  Part of this work was undertaken on the Cambridge CSD3 part of the STFC DiRAC HPC Facility (www.dirac.ac.uk) funded by BEIS capital funding via STFC Capital Grants ST/P002307/1 and ST/R002452/1 and STFC Operations Grant ST/R00689X/1.

\appendix

\section{Joint Analysis of Bispectrum Shapes}
\label{app:joint}

We review the statistical formalism for constraining multiple bispectrum shapes and provide some details on how Figure \ref{fig:ellipse} was made.

\subsection{Model}

Suppose that we have a model of the early universe in which the primordial bispectrum can be parametrized as
\begin{align}
    B_{\zeta\zeta\zeta}(k_1,k_2,k_3) = \sum_{i=1}^{n} \fNL^{(i)} B^{(i)}(k_1,k_2,k_3),
\end{align}
for $n$ template bispectrum shapes $B^{(1)},\cdots,B^{(n)}$. It's common to place a normalization condition to fix the amplitude of these templates at $(k_1,k_2,k_3)=(k_*,k_*,k_*)$ with $k_*=0.05 \mathrm{Mpc}^{-1}$.

We would like to test this model using the CMB. Our statistical model for the late-time CMB bispectrum constructed from observations is then:
\begin{align}
    b_\ells^\mathrm{obs} = \sum_{i=1}^{n} \fNL^{(i)} b^{(i)}_\ells+ \epsilon_\ells,
\end{align}
for $\ell_1 \le \ell_2 \le \ell_3$. The noise term $n_\ells$ comes from both cosmic variance and measurement noise. Our model is still linear in $\fNL$s since the primary CMB anisotropies relate linearly to $\zeta$.

In vector form,
\begin{align}
    \mathbf{b}^\mathrm{obs} &= \sum_{i=1}^{n} \fNL^{(i)} \mathbf{b}^{(i)} + \boldsymbol{\epsilon} \\
    &= X \mathbf{f} + \boldsymbol{\epsilon},
\end{align}
where $(\mathbf{f})_i=\fNL^{(i)}$ and $X_{\boldsymbol{\ell}i}=(\mathbf{b}^{(i)})_{\boldsymbol{\ell}}$. We assume (near-)Gaussian errors on this bispectrum statistic: $\boldsymbol{\epsilon} \sim \mathcal{N}(\mathbf{0}, \Sigma)$.  \footnote{The covariance $\sigma$ is often assumed to be diagonal for computational convenience. $\Sigma_{\ells, \ell'_1 \ell'_2 \ell'_3} = \delta_{\ell_1 \ell'_1} \delta_{\ell_2 \ell'_2} \delta_{\ell_3 \ell'_3} C_{\ell_1} C_{\ell_2} C_{\ell_3}$ up to a symmetry factor. The $C_\ell$ denotes the CMB power spectrum including instrumental noise. }

\subsection{Estimator}

The likelihood of $\fNL$s given observation up to some constant (independent of $\fNL$s) factors is given by:
\begin{align}
    \mathcal{L}(\mathbf{f} | \mathbf{b}^\mathrm{obs}) &= \mathrm{Prob}\left[ \boldsymbol{\epsilon} =\left(\mathbf{b}^\mathrm{obs} - X\mathbf{f} \right) \right] \\
    &\propto \exp \left[ - \frac{1}{2} \left( \mathbf{b}^\mathrm{obs} - X\mathbf{f} \right)^\top \Sigma^{-1} \left( \mathbf{b}^\mathrm{obs} - X\mathbf{f} \right) \right] \\
    &\propto \exp \left[ - \frac{1}{2} \left( \mathbf{f} - \hat{\mathbf{f}} \right)^\top F \left( \mathbf{f} - \hat{\mathbf{f}} \right) \right], \label{eqn:likelihood}
\end{align}
where
\begin{align}
    \hat{\mathbf{f}} &\equiv \left( X^\top \Sigma^{-1} X \right)^{-1} X^\top \Sigma^{-1} \mathbf{b}^\mathrm{obs}, \\ 
    F &\equiv  X^\top \Sigma^{-1} X.
\end{align}
The likelihood \eqref{eqn:likelihood} is again multivariate normal: $\mathbf{f} \sim \mathcal{N}(\hat{\mathbf{f}},F^{-1})$. This gives our constraint on the $\fNL$s given the CMB data. The likelihood is maximised at $\mathbf{f}=\hat{\mathbf{f}}$, which is our optimal\footnote{The estimator is unbiased and has the smallest variance among unbiased estimators.} estimate of the $\fNL$s given the data.

The matrix $F$ is the Fisher information matrix. The more information we have about $\mathbf{f}$, the larger the $F$, and the tighter error bars $F^{-1}$ on the $\fNL$s. One can check that this corresponds to the usual bispectrum estimator when $n=1$. 

Note that
\begin{align}
    F_{ij} = \left( \mathbf{b}^{(i)} \right)^\top \Sigma^{-1}  \mathbf{b}^{(j)}.
\end{align}
This can be interpreted as an inner product between the $i$th and $j$th templates, with weights given by $\Sigma^{-1}$. As long as the templates are not degenerate under this inner product, $F$ is a positive definite matrix (has real, positive eigenvalues).

\subsection{Constraints}

Given no prior knowledge of the $\fNL$s, the likelihood \eqref{eqn:likelihood} provides the constraints on $\fNL$ given the data. In Bayesian terms, this is our \textit{posterior} distribution for a uniform prior without bounds.

In the simple case of $n=1$, we have only one template and $\mathbf{f}=\fNL$, so the constraint can be written simply as
\begin{align}
    \fNL = \hat{f} \pm F^{-1/2} \quad \quad \mathrm{at\ 68\%\ CL}.
\end{align}
Our best-fit $\fNL$ value is given by $\hat{f}$ and the $1\sigma$ error by $1/\sqrt{F}$. 

Otherwise, we have a multivariate normal distribution: $\mathrm{P}\left[ \mathbf{f} | \mathrm{data} \right] \sim \mathcal{N}(\hat{\mathbf{f}},F^{-1})$. This provides a \textit{joint} constraint to the $\fNL$s of multiple bispectrum templates. The contours for the $\fNL$s are centered around $\hat{\mathbf{f}}$, the best-fit value of the $\fNL$s. 

The shape of the contours depends on $F^{-1}$. $F^{-1}$ is symmetric and positive definite and hence has real eigenvalues and eigenvectors. Suppose that $F^{-1}$ has eigenvectors $\mathbf{v}_1,\cdots,\mathbf{v}_n$ with corresponding eigenvalues $\lambda_1,\cdots,\lambda_n >0$. Then the $1\sigma$ contour, for example, has the shape of an ellipsoid with principal axes along $\mathbf{v}_1,\cdots,\mathbf{v}_n$ with the semi-axes length given by $\sqrt{\lambda_1}, \cdots, \sqrt{\lambda_n}$. Its volume is equal to  $\pi\sqrt{\lambda_1}\cdots\sqrt{\lambda_n} = \sqrt{\mathrm{det}F^{-1}}$. \footnote{To see this, consider $\mathbf{x} \equiv P^{-1} (\mathbf{f}-\hat{\mathbf{f}})$, where $P=(\mathbf{v}_1 | \cdots | \mathbf{v}_n)$. $\mathbf{x}$ has vanishing mean, and its variance is given by $P^{-1} F^{-1} (P^{-1})^\top = \mathrm{diag}\{\lambda_1, \cdots, \lambda_n \} $ (assuming that the eigenvectors are normalised to unity and $P^T=P^{-1}$). So the $1\sigma$ contour for $(x_1/\sqrt{\lambda}_1,\cdots,x_n/\sqrt{\lambda}_n )$ is the unit sphere. }

If the bispectrum templates are orthogonal to each other in the late-time space, $F$ is diagonal. In this case, $F^{-1}=\mathrm{diag}\{ 1/F_{11}, \cdots, 1/F_{nn} \}$, and the constraints on each $\fNL$ are independent from each other. The posterior is a product of $n$ one-dimensional Gaussian distributions.

\vspace{10pt}

We can obtain \textit{marginalized} constraints by integrating out the parameters we are not interested in. For example, marginalizing over all $\fNL$s except the first gives
\begin{align}
    \mathrm{P}\left[ \fNL^{(1)} | \mathrm{data} \right] &= \int d\,\fNL^{(2)} \cdots d\,\fNL^{(n)}  \ \mathrm{P}\left[ \mathbf{f} | \mathrm{data} \right]   \\
    &\sim \mathcal{N}\left( (\hat{\mathbf{f}})_1,\left(F^{-1}\right)_{11} \right).
\end{align}
This is again a Gaussian distribution: a 1D projection of the ellipsoid in $n$ dimensions. Note that $\left(F^{-1}\right)_{11} \geq \left( F_{11} \right)^{-1}$ in general; we obtain looser constraints by considering a more general model. If the bispectrum template $\mathbf{b}^{(1)}$ is orthogonal to the others, however, the equality holds.

\vspace{10pt}
We can obtain \textit{conditional} constraints by setting some of the $\fnl$s to a fixed value. For example, with $n=2$ and $\fNL^{(2)}=\alpha$ fixed, the conditional distribution of $\fNL^{(1)}$ is given by
\begin{align}
    \mathrm{P}\left[ \fNL^{(1)}\ |\ \mathrm{data},\ \fNL^{(2)}=\alpha \right] \sim \mathcal{N}\left( f'(\alpha) , \sigma^2 \right),
\end{align}
where
\begin{align}
    f'(\alpha) &\equiv (\hat{\mathbf{f}})_1 - \left(F^{-1}\right)_{12} (\alpha-(\hat{\mathbf{f})_2})/\left(F^{-1}\right)_{22} \\
    \sigma^2 &\equiv \left(F^{-1}\right)_{11} - \left(F^{-1}\right)_{12} \left(F^{-1}\right)_{21} / \left(F^{-1}\right)_{22}.
\end{align}
If the two shapes are orthogonal, then $F$ is diagonal and $(F^{-1})_{12}=(F^{-1})_{21}=0$, and we recover the case for $n=1$. Conditional distributions are identical to the independent ones for orthogonal shapes.

\subsection{Example - Figure \ref{fig:ellipse}}

For the example shown in Figure \ref{fig:ellipse}, we set the Fisher information as
\begin{align}
    F &= \begin{pmatrix} \sigma_1^{-2} & \rho \sigma_1^{-1} \sigma_2^{-1} \\ \rho \sigma_1^{-1} \sigma_2^{-1} & \sigma_2^{-2} \end{pmatrix} \\ 
    &= \left[ \frac{1}{1-\rho^2} \begin{pmatrix} \sigma_1^{2} & -\rho \sigma_1 \sigma_2 \\ -\rho \sigma_1 \sigma_2 & \sigma_2^{2}  \end{pmatrix} \right]^{-1} \equiv N^{-1}.
\end{align}
We vary the correlation coefficient $\rho$ as 0, 0.6, and 0.99, while keeping the best-fit values $\hat{\mathbf{f}}=(0.5,0)$, for the left plot of Figure \ref{fig:ellipse}.

The resulting estimates are bivariate normal: $\mathbf{f}\sim\mathcal{N}(\hat{\mathbf{f}},N)$. The $1\sigma$ and $2\sigma$ contours are given by the iso-contours $(\mathbf{f}-\hat{\mathbf{f}})^\top N^{-1}(\mathbf{f}-\hat{\mathbf{f}})=$ 1 and 2, respectively. The two principal axes and their lengths are given by the eigenvectors and their eigenvalues of $F$. The marginalized constraints: $\fnl^{(i)}=\hat{\mathbf{f}}_i \pm \sigma_i/\sqrt{1-\rho^{2}}$ (68\% CL). 

Next, consider orthogonalizing the second shape, so that $S=f_\mathrm{NL}^{\ (1)} S^{(1)} + f_\mathrm{NL}^{\ (2)} S^{(2)}$ becomes $S=f_\mathrm{NL}^{\ (1)} S^{(1)} + \tilde{f}_\mathrm{NL}^{\ (2)}(S^{(2)}-xS^{(1)})$. This is equivalent to a redefinition
\begin{align}
    \begin{pmatrix} f_\mathrm{NL}^{\ (1)} \\ f_\mathrm{NL}^{\ (2)} \end{pmatrix} = \begin{pmatrix} 1 & -x \\ 0 & 1 \end{pmatrix} \begin{pmatrix} \tilde{f}_\mathrm{NL}^{\ (1)} \\ \tilde{f}_\mathrm{NL}^{\ (2)} \end{pmatrix}.
\end{align}
It can be shown that $x=F_{01}/F_{11}$ orthogonalizes the shapes, which gives the right-hand plot of Figure \ref{fig:ellipse}.

\bibliographystyle{utphys}
\bibliography{references.bib}
\end{document}